\documentclass[sigconf, nonacm]{acmart}

\usepackage{diagbox}
\usepackage{amsmath} 
\usepackage{adjustbox}
\usepackage{subcaption}
\usepackage{float}  
\usepackage{pdflscape} 
\usepackage{rotating}  
\usepackage{stmaryrd}  
\usepackage{tikz}
\usetikzlibrary{shadows, shapes.geometric, fit, positioning}
\usepackage{pgfplots} 
\pgfplotsset{compat=1.13}
\usepackage{apxproof} 
\usepackage{mathpartir} 
\usepackage{balance}
\usepackage{tabularx,array,xparse,makecell,etoolbox}
\usepackage[linesnumbered,ruled,vlined]{algorithm2e} 
\usepackage[disable]{todonotes} 
\usepackage[utf8]{inputenc}
\usepackage{caption}
\newbox\boiteamu
\sbox\boiteamu{\large \itshape \bfseries \textmu}

\newcommand{\ruleref}[1]{\textcolor{green!60!black}{\ref{#1}}}
\newcommand{\costref}[1]{\textcolor{red}{\ref{#1}}}

\newtheorem{defi}{Definition}

\newtheoremrep{thm}{Theorem}
\newtheoremrep{prop}{Proposition}
\newtheoremrep{ppty}{Property}
\newtheoremrep{lem}{Lemma}

\newcommand{\ra}{\textsc{A3D-RA}}

\newcommand{\aj}{\texttt{arrayJoin }}

\newcommand{\halonly}[1]{}

\newcolumntype{Y}{>{\centering\arraybackslash}X}

\newcommand{\sem}[1]{{\llbracket#1\rrbracket}}

\newcommand{\filtN}[1][\theta]{\sigma_{#1}}
\newcommand{\filt}[2][\theta]{\sigma_{#1}(#2)}
\newcommand{\proj}[2]{\pi_{#1}(#2)}
\newcommand{\union}[0]{ \cup }
\newcommand{\join}{\bowtie}
\newcommand{\arrfiltN}[1][a,\theta]{\phi_{#1}} 
\newcommand{\arrfilt}[2][a:n,\theta]{\phi_{#1}(#2)}
\newcommand{\arrfilts}[2][(a_1:n_1,\theta_1),\dots,(a_k:n_k,\theta_k)]{\phi_{#1}(#2)}

\newcommand{\arrjoinN}[1][a]{\mu_{#1}}
\newcommand{\arrjoin}[2]{\mu_{#1}(#2)}
\newcommand{\arrjoins}[4][k]{\mu_{{#2}_1:{#3}_1,\dots,{#2}_{#1}:{#3}_{#1}}(#4)}
\newcommand{\deriveN}[1][y=f(x_1,\dots,x_s)]{\delta_{#1}}
\newcommand{\derive}[2][y=f(x_1,\dots,x_s)]{\delta_{#1}(#2)}

\newcommand{\aggN}[1][G,f(x_1,\dots,x_s):n]{\Gamma_{#1}}
\newcommand{\agg}[2][G,f(x_1,\dots,x_s):n]{\Gamma_{#1}(#2)}
\newcommand{\aggs}[2][G,f_1(x^{1}_{1},\dots,x^{1}_{s1}):n_1,\dots,f_k(x^{k}_{1},\dots,x^{k}_{sk}):n_k]{\Gamma_{#1}(#2)}

\NewDocumentCommand{\aggf}{o}{%
  \IfNoValueTF{#1}
    {\fct{\texttt{Agg}}}%
    {\fct{\texttt{Agg}}{#1}}%
    }%

\NewDocumentCommand{\aggfirst}{o}{%
  \IfNoValueTF{#1}
    {\fct{\texttt{Agg}^{(i)}}}%
    {\fct{\texttt{Agg}^{(i)}}{#1}}%
    }%
\NewDocumentCommand{\aggfin}{o}{%
  \IfNoValueTF{#1}
    {\fct{\texttt{Agg}^{(f)}}}%
    {\fct{\texttt{Agg}^{(f)}}{#1}}%
  }%

\NewDocumentCommand{\fct}{m g}{%
  \ensuremath{%
    {#1}%
    \IfNoValueTF{#2}{}{(#2)}%
  }%
}
\newcommand{\bag}[1]{\{\!\{#1\}\!\}}

\newcommand{\aggforeach}[2][]{\texttt{AggForEach}_{#1}(#2)}

\newcommand{\eq}{\equiv}

\newcounter{queryCounter}

\newcommand{\NJoin}{\bowtie}

\newcolumntype{C}[1]{>{\centering\arraybackslash}p{#1}}
  
\newenvironment{packed-itemize}
{\leftmargini=15pt\begin{itemize}
\setlength{\itemsep}{-0.3pt}}
{\end{itemize}\vspace{-0.1em}}

\newenvironment{packed_inner-itemize}
{\begin{packed-itemize}}
{\end{packed-itemize}\vspace{0.25em}}

\newcounter{rule}
\renewcommand{\therule}{R\arabic{rule}}
\newcommand{\ruleeq}[2][]{%
  \refstepcounter{rule}
  \begin{equation}
    \tag{\therule#1}
    \label{\therule#1}%
    #2
  \end{equation}
}

 \newcounter{subrule}[rule]
 
\newcommand{\subruleeq}[2][]{%
  \refstepcounter{subrule}%
  \begin{equation}
    \tag{\therule#1}
    \label{\therule#1}
    #2
  \end{equation}
 }

 \newcommand{\vsqueezeabovecaption}{\vspace{-0.22cm}}

\newcommand{ \vsqueezeaftercaption}{\vspace{-0.3cm}}

\begin{document}

\fancyhead{} 

\title{Optimizing Relational Queries over Array-Valued Data in Columnar Systems}

\author{Maroua Zeblah}
\affiliation{%
  \institution{Tyrex team, Univ. Grenoble Alpes, CNRS, Inria \& Core Engine team, Opensee, Paris}
}
\email{maroua.zeblah@inria.fr}

\author{Etienne Couritas}
\affiliation{%
  \institution{Core Engine team, Opensee, Paris}
}
\email{etienne.couritas@opensee.io}

\author{Sarah Chlyah}
\affiliation{%
  \institution{Tyrex team, Univ. Grenoble Alpes, CNRS, Inria, Grenoble INP, LIG, 38000 Grenoble, France}
}
\email{}

\author{Pierre Genev\`es}
\affiliation{%
  \institution{Tyrex team, Univ. Grenoble Alpes, CNRS, Inria, Grenoble INP, LIG, 38000 Grenoble, France}
}
\email{}

\author{Nils Gesbert}
\affiliation{%
  \institution{Tyrex team, Univ. Grenoble Alpes, CNRS, Inria, Grenoble INP, LIG, 38000 Grenoble, France}
}
\email{}

\author{Nabil Laya\"ida}
\affiliation{%
  \institution{Tyrex team, Univ. Grenoble Alpes, CNRS, Inria, Grenoble INP, LIG, 38000 Grenoble, France}
}
\email{}

\begin{abstract}
Modern analytical workloads increasingly combine relational data with array-valued attributes. While columnar database systems efficiently process such workloads, their ability to optimize queries that interleave relational operators with array manipulations remains limited. This paper introduces A3D-RA, an extended relational algebra supporting array-valued attributes, together with a comprehensive framework for algebraic reasoning and optimization. We formalize its data model and semantics, develop a complete set of equivalence-preserving transformation rules capturing pairwise interactions between relational and array operators, and propose a plan enumeration strategy with an optimality guarantee that remains polynomial in all non-join operators. We design A3D-RA as a modular, backend-independent optimization layer that can be instantiated over existing analytical database systems. Experimental results across three high-performance engines on a real-world workload show consistent performance gains enabled by the proposed algebraic optimization layer.
\end{abstract}

\maketitle

\vspace{2.5cm}

\section{Introduction}
Modern analytical workloads increasingly involve array-valued data, such as time series in financial applications, feature vectors in machine learning pipelines, or sensor readings in IoT systems.
Such data naturally combines relational structure (e.g., trades, customers, devices) with multi-dimensional arrays stored as attributes.
It is often stored in a denormalized form~\cite{jaeschke-pods82} and processed using column-oriented database engines such as Amazon Redshift~\cite{redshift-sigmod2015}, Snowflake~\cite{snowflake2016}, Apache Pinot~\cite{pinot-sigmod18}, DuckDB~\cite{duckdb-sigmod2019}, Umbra~\cite{neumann-umbra2020}, and ClickHouse~\cite{schulze2024clickhouse}. 
Column stores are particularly well suited to read-heavy analytical workloads of this kind, as they support efficient compression~\cite{abadi-sigmod2006} and vectorized execution~\cite{monetdb2005}.
While denormalization and array-valued attributes improve data locality and simplify modeling, they also introduce fundamental challenges for query processing and optimization.
Many existing database engines support the evaluation of array operations through proprietary extensions or user-defined functions~\cite{rusu-ftdb2023}, but these mechanisms often lack a unified algebraic foundation, leaving room for further optimization opportunities of high-level analytical queries. 

To address this gap, we propose A3D-RA, a formal and modular algebraic optimization framework for processing queries that combine relational and array operations.
Rather than targeting a specific execution engine, A3D-RA is designed as a DBMS-agnostic logical optimization layer that can be instantiated on top of existing analytical systems supporting array-valued attributes.

The goal is to extend the expressive power of relational algebra to efficiently handle array-valued attributes commonly used in analytical systems, thus enabling a systematic and global exploration of operator interactions.

\paragraph{Contributions.}
Specifically, this paper makes the following contributions:
(i) we introduce an \emph{extended relational algebra} (A3D-RA) that supports array-valued attributes, defining its data model together with the syntax and formal semantics of its operators; (ii) we propose a comprehensive and compositional set of equivalence-preserving transformation rules capturing the interactions between all pairs of relational and array operators, enabling algebraic reasoning and systematic plan rewriting in the presence of arrays; (iii) we present a plan enumeration strategy with an optimality guarantee that is polynomial in all non-join operators; and finally (iv) 
we design A3D-RA as a modular optimization layer that can be instantiated over different database engines. We denote by A3DRA[X] the instantiation of our framework over a backend system X. We implement a prototype with a pluggable backend architecture and evaluate three instantiations (A3DRA[ClickHouse], A3DRA[Umbra], and A3DRA[Snowflake]) on  queries derived from a real-world financial workload. The results demonstrate that our algebraic framework consistently unlocks optimization opportunities beyond those explored by native optimizers, without requiring modifications to the underlying execution engines.

\section{the A3D-extended rel.\ algebra}
\label{sec:mus}

\subsection{Data model}
The considered data model extends the classical relational algebra to efficiently handle denormalized and multidimensional data. In this model, a \emph{relation} is defined as a set of \emph{tuples}, where each tuple (also called mapping or row) maps \emph{column names} to \emph{values}. Unlike traditional relational models, our approach supports columns that can hold \emph{arrays of values}, enabling the representation of hierarchical and multidimensional data in a compact and flexible form.

Formally, we define the following sets:

\begin{itemize}
\item $\mathfrak{V}$ an infinite set of \emph{values}, including scalars and arrays
\item $\mathfrak{C}$ an infinite set of \emph{column names}
\end{itemize}

\begin{defi}[Tuple]\label{def:tuple}
A \emph{tuple} is a partial function $t\colon\mathfrak{C}\to
\mathfrak{V}$ whose domain is finite. If $dom(t) =
\{c_1,\ldots,c_n\}$, $t$ can also be seen as the set
$\{c_1\rightarrow t(c_1), \ldots, c_n\rightarrow t(c_n) \}$.
\end{defi}


\begin{defi}[Relation]\label{def:relation}
Let $\mathcal{C}$ be a finite set of \emph{column names}. A \emph{relation} $R$ of \emph{type} $\mathcal{C}$ is a finite set of \emph{tuples}
$t : \mathcal{C} \to \mathfrak{V}$.
We partition $\mathcal{C}$ into disjoint subsets:
$
  S \subseteq \mathcal{C} (\text{scalar columns}),
  A \subseteq \mathcal{C} (\text{array columns}),
$
such that $\mathcal{C} = S \cup A$ and $S \cap A = \varnothing$.
Each tuple $t \in R$ then satisfies:
\begin{enumerate}
  \item For every scalar column $c \in S$, $
      t(c) \in V_{\mathrm{scalar}}$.

  \item For every array column $c \in A$, $t(c) \in V_{\mathrm{array}} 
      = \mathcal{M}(V_{\mathrm{scalar}})$, 
        where $\mathcal{M}(V_{\mathrm{scalar}})$ denotes the set of all finite sequences of values from $V_{\mathrm{scalar}}$.

\end{enumerate}
Hence, equivalently,
$
  R \subseteq
  \displaystyle \prod_{c \in S} V_{\mathrm{scalar}} \;\times\;
  \prod_{c \in A} V_{\mathrm{array}},
$ subject to $|R| < \infty$.
\end{defi}

\begin{defi}[Semantic Correspondence of Arrays]\label{def:corr}
Assume a binary \emph{semantic correspondence} relation 
$
\sim \;\subseteq\; V \times V
$
over a universe of values \(V\). 
Two arrays 
$
A = [a_1, a_2, \dots, a_n]
\text{ and }
B = [b_1, b_2, \dots, b_n]
$
are said to be \emph{semantically corresponding} (denoted \(A \approx B\)) if and only if:

\begin{enumerate}
  \item They have the same length \(n\), and
  \item For every index \(i\in\{1,\dots,n\}\),  $a_i \sim b_i$.
  \end{enumerate}
That is,
$
A \approx B \;\Leftrightarrow \;
\bigl|A\bigr| = \bigl|B\bigr|
\;\wedge\;
\forall\,i=1,\dots,n,\;a_i \sim b_i.
$
Each pair \((a_i, b_i)\) thus stands in the semantic correspondence relation.
\end{defi}

\subsection{Syntax and Semantics of A3D-RA terms} 
The \ra{} algebra extends the traditional relational algebra by incorporating  operators for transforming and computing over relations that contain array columns. 

\subsubsection{Terms}\label{sec:terms} The core syntax of terms is defined in
Fig.~\ref{fig:syntax}. Base terms are relation variables $R$. The first three operators are standard from classical relational algebra: 
two relations can be joined with the \emph{natural join}
operator $\Join$; one relation can be filtered using the selection operation $\sigma_{\theta}$ where $\theta$ is a filter (further detailed below); and the projection operator $\proj{L}{\varphi}$ selects the desired columns $L$ in  $\varphi$ while discarding other columns.  
The three operators \textit{ArrayJoin}, \textit{Array-filter}, and \textit{Aggregation} enable transformations and computations over array-valued columns, while  \textit{Derive} creates new columns.
The formal semantics of all operators, including their behavior over array-typed attributes, are given in Fig.~\ref{fig:sem}.

\begin{figure}
  \begin{small}
  \centering
  \begin{tabular}{cclr}%
    $\varphi$ & $::=$ & ~ & term\\
    &&  $R$ &  relation variable \\
    & $|$ &  $\varphi_1 \join \varphi_2$ & natural join \\
    & $|$ &  $\filt{\varphi}$ & filtering \\
     & $|$ &  $\proj{L}{\varphi}$ & projection \\
    & $|$ &  $\arrfilt{\varphi}$ & array-filter \\
    & $|$ &  $\arrjoin{a:n}{\varphi}$ & arrayJoin\\
    & $|$ &  $\derive{\varphi}$ & derive \\
    & $|$ &  $\agg{\varphi}$ & aggregate \\
    
  \end{tabular}\\
  \caption{Syntax of {\ra{}} terms.}
  \label{fig:syntax}
  
\end{small}
\end{figure}

  \begin{figure*}
    \centering\small
       \begin{adjustbox}{width=\textwidth,center}
      \renewcommand\arraystretch{1.4}\renewcommand\dots{...}
      $
      \begin{array}{rclcrcl}

  \sem{\varphi_1  \NJoin  \varphi_2} &=& \{ t_1 \cup t_2  \mid  t_1\in\sem{\varphi_1} \land t_2\in\sem{\varphi_2} \land  \forall c \in dom(t_1) \cap dom(t_2)~ t_1(c) = t_2(c) \}
    \\
  \sem{\filt{\varphi}}& = & \{ t \mid t\in\sem{\varphi} \land \mathfrak \theta(t)=\top \} 
 \\
   \sem{\proj{L}{\varphi}}  & = & \Bigl\{ \{ c \rightarrow v \in t  \mid   c\in L\} \Bigm| t\in \sem{\varphi} \Bigr\}  
   \\
  \sem{\arrfilt{\varphi}} & = & \Bigl\{ \bigl\{ c \rightarrow v \in t \bigm| c \notin \{a, n\}\bigr\} \cup \bigl\{n \rightarrow \{e \mid e \in v \land \theta(e)\}  \bigm|   a \rightarrow v \in t\bigr\} \Bigm| t \in \sem{\varphi} \Bigr\} 
   %
    \\
   \sem{\arrfilts[(a_1:n_1\dots a_k:n_k,\theta)]{\varphi}} & = & \Bigl\{ \bigl\{ c \rightarrow v \in t \bigm| c \notin \{a_1, n_1\dots a_k, n_k\} \bigr\} \cup \bigl\{n_i \rightarrow \{e_i^j \mid 1\leqslant j \leqslant s\land \theta(e_{1}^j\dots e_k^j)\} \bigm|  1\leqslant i \leqslant k \land a_i \rightarrow [e_i^1\dots e_i^s] \in t\bigr\} \Bigm| t \in \sem{\varphi} \Bigr\} 
   \\
    \sem{\arrjoins{a}{n}{\varphi}}  & = & 
    \Bigl\{ \bigl\{ c \rightarrow v \in t \bigm| c \notin \{a_1, n_1\dots a_k, n_k\} \bigr\} 
      \cup \bigl\{n_i \rightarrow e_i^j \bigm| 
       1\leqslant i\leqslant k \land
      a_i \rightarrow [e_i^1\dots e_i^s] \in t\bigr\} \Bigm|
    t\in \sem{\varphi} \land 
      1\leqslant j\leqslant s\land
     s=|t(a_i)|
    \Bigr\}
   \\
      \sem{\derive{\varphi}}  & = & \Bigl\{ \bigl\{c \rightarrow v \in t \bigm| c \notin y \bigr\} \cup \bigl\{y \rightarrow f(v_1\dots v_{s}) \bigm| 
  1\leqslant j\leqslant s\land x_{j} \rightarrow v_j \in t\bigr\} \Bigm| t \in \sem{\varphi} \Bigr\} 
%
%
    \\
  \sem{\aggs[G,f_1(x^{1}_{1}\dots x^{1}_{s_1}):n_1\dots f_k(x^{k}_{1}\dots x^{k}_{s_k}):n_k]{\varphi}}  & = & 
    \Bigl\{ y \cup \bigl\{n_i \rightarrow a_i \bigm|  1\leqslant i\leqslant k 
    \land a_i = f(S_i) \land 
    S_i = \{(v_1\dots v_{s_i})\mid t \in \sem{\varphi} \land t.G = y.G 
    \land \forall j \in \{1\dots s_i\},x^{i}_{j} \rightarrow v_j \in t \} \bigr\}
  \Bigm|
  y \in \sem{\proj{G}{\varphi}}
    \Bigr\}
    \end{array}
    $
  \end{adjustbox}
  \caption{Semantics of \ra{}.}
  \label{fig:sem}
  \end{figure*}

\subsubsection{Filters} The standard filter operation
$\sigma_\theta$, which operates on a relation by keeping only a
subset of its tuples, depends on a \emph{predicate} $\theta$
indicating which tuples are to be kept. This predicate can be seen as a
function from mappings to booleans. The predicate can be any legally formed expression that involves:
constants (i.e., members of any column domain); 
column names (i.e., a subset of the columns from the expression on which the filter is applied); arithmetic comparisons (=, $\neq$ , <, $\leqslant$, >, $\geqslant$); and logical operators (and, or, not).





\begin{defi}[Invertible Predicate] \label{invertible-filter}
Let $\theta$ be a predicate of the form $x \mapsto R(x, v)$ where $v$ is a constant, $R$ a binary relation (e.g., $=$, $>$, $\leq$). Let $f$ be a transformation function.
The predicate $\theta \circ f : x \mapsto R(f(x), v)$ is said to be \emph{invertible} if there exist a relation $R'$ and a function $f'$ such that: $R(f(x), v) \equiv \theta'(x)$ where $\theta' : x \mapsto R'(x, f'(v))$. Notice that $f'(v)$ is a constant so that $\theta'$ avoids computing $f$ while yielding the same result as  $\theta \circ f$. \\
For example, $3-x<15$ can be rewritten into $x>-12$. 
\end{defi}

\subsubsection{Array-Join Operator $\mu$} \mbox{}\\[0.3ex]
\label{sec:arrayjoin}
The \texttt{arrayJoin} operator $\arrjoin{a:n}{\varphi}$ unnests the array column $a$, creating one row per element while duplicating the other columns of $\varphi$. This expansion increases the number of rows proportionally to the array size, producing the \textit{flattened} columns. 
For example:
\begin{center}
\begin{footnotesize}
\renewcommand{\arraystretch}{0.9}
\setlength{\tabcolsep}{3pt}
\begin{tabular}{|c|c|}
\hline
\textbf{id} & \textbf{vals} \\
\hline
1 & [10, 20, 30] \\
2 & [5, 15] \\
\hline
\end{tabular}
\raisebox{0.1cm}{$\xrightarrow[\text{}]{\texttt{arrayJoin(vals)}}$}
\begin{tabular}{|c|c|}
\hline
\textbf{id} & \textbf{vals} \\
\hline
1 & 10 \\
1 & 20 \\
1 & 30 \\
2 & 5 \\
2 & 15 \\
\hline
\end{tabular}
\end{footnotesize}
\end{center}

Formally, its interpretation is:
%
$$\sem{\arrjoin{a:n}{\varphi}}   =  
  \Big\{\{ c \rightarrow v \in t \mid c \neq a \}  \cup 
\{n \rightarrow e\}
~\Big|~ e \in t(a) \land t \in \sem{\varphi} \Big\} $$
where subscripts indicate that $a$ is the array column to be flattened; and $n$ (optional) denotes the new name for the scalar column resulting from the array. 
If not specified, the original name $a$ is reused.
This convention consistently applies to all operators generating new or transformed columns. 
As shown in Fig.~\ref{fig:sem}, the $\aj$ operator can simultaneously unnest multiple array columns of equal length per row, ensuring element-wise alignment across arrays.




\subsubsection{Array-Filter Operator $\phi$}\mbox{}\\[0.3ex]
\label{sec:arrayfilter}
The \texttt{arrayFilter} operator $\arrfilt{\varphi}$ applies a predicate $\theta$ to the elements of an array column $a$, returning a new array that contains only the elements satisfying $\theta$. The number of rows in $\varphi$ is preserved, although some arrays may become empty.  
This operator is also extended to multiple corresponding array columns, denoted as $\arrfilts[(a_1:n_1,\dots,a_k:n_k,\theta)]{\varphi}$, ensuring that positional correspondence between arrays is maintained after filtering (i.e., the filter is applied across all of them in a coordinated way).

\subsubsection{Derive Operator $\delta$}\mbox{}\\[0.3ex]
\label{sec:derive}
The \texttt{derive} operator $\derive{\varphi}$ extends a relation $\varphi$ with a new column $y$, whose values are obtained by applying a function $f$ to one or more existing columns $(x_1,\dots,x_s)$. The function $f$ may represent arithmetic, string, or other domain-specific transformations.  We assume that its computational cost is linear with respect to the total length of the input arrays. 
When applied to array columns, the operator relies on the $\texttt{arrayMap}_{(f)}$ construct, where $f$ is applied element-wise to the arrays provided as parameters.

\subsubsection{Aggregation Operator $\Gamma$}\mbox{}\\[0.3ex]
\label{sec:agg}
The \texttt{aggregation} operator $\agg{\varphi}$ groups the rows of a relation $\varphi$ according to a set of attributes $G$ and computes aggregate values using a function $f$ over columns $(x_1, \dots, x_k)$.  
If $G$ is empty, the entire relation is treated as a single group. The operator produces one row per group, with the result of $f$ stored in a column $n$ (defaulting to $f(x_1,\dots,x_k)$).  %
Multiple aggregations can be computed simultaneously, denoted by $\aggs{\varphi}$.  

\section{Generating new query plans} 
\label{sec:rules}
We present the equivalence-preserving transformation rules for \ra{}, derived from a systematic analysis of pairwise operator interactions (Table~1). Rules are classified as \emph{rule-based}, when universally beneficial regardless of data statistics or implementation details, and \emph{cost-based}, when their application depends on estimated execution cost.

\subsection{Aggregation-free rewrite rules}  
\label{newrules}

\begin{table*}
  \begin{center}
   \begin{small}
  \renewcommand\arraystretch{1.1}
\[  
\begin{array}{|C{2.3cm}|C{2.3cm}|C{2.3cm}|C{2.3cm}|C{2.3cm}|C{2.3cm}|C{2.3cm}|}
\hline
\textbf{Operators} & \textbf{Filter ($\sigma$)} & \textbf{Proj ($\Pi$)} & \textbf{ArrayJoin ($\mu$)} & \textbf{ArrayFilter ($\phi$)} & \textbf{Derive ($\delta$)}  & \textbf{Join ($\bowtie$)}  \\
\hline
\textbf{ArrayJoin ($\mu$)} & \makecell{\ruleref{R2.1} \\ \ruleref{R2.2} \costref{R2.3}}  &
\ruleref{R3}& \costref{R1} & \ruleref{R6} & \makecell{\ruleref{R5.1}  \costref{R5.2}} & \makecell{\costref{R4.1}  \costref{R4.2}}  \\
\hline
\textbf{ArrayFilter ($\phi$)}&\makecell{\ruleref{R8}}   &
\ruleref{R9}& - &\costref{R7}  & \makecell{\costref{R11.1} \ruleref{R11.2} } & \makecell{\costref{R10.1}  \costref{R10.2} \\ \costref{R10.3}} \\
\hline
\textbf{Derive ($\delta$)} & \makecell{\ruleref{R13.1} \\ \ruleref{R13.2}}   &
\ruleref{R14}& - & - & \costref{R12}  & \makecell{\costref{R15}} \\
\hline
\textbf{Aggregate ($\Gamma$)}& \makecell{\costref{R16} \\ \costref{R18}} & - &\makecell{\costref{R17.1}  \costref{R17.2} \\ \costref{R17.3}} & \costref{R20} & \costref{R19} & \costref{R21}  \\
\hline
\end{array}\]
  \begin{tabular}{llllll}
    \textcolor{green!60!black}{\rule{1ex}{1ex}} & Rule-based transformations &
    \textcolor{red}{\rule{1ex}{1ex}} & Cost-based transformations &
  \end{tabular}
  \caption{Rewrite Rules of {\ra}.}
  \label{fig:rules}
\end{small}
\end{center}
\vsqueezeaftercaption{}
\vsqueezeaftercaption{}
\end{table*}

\subsubsection{Commutativity of ArrayJoin}
\ruleeq{
\arrjoin{a_2:n_2}{\arrjoin{a_1:n_1}{\varphi}}\eq \arrjoin{a_1:n_1}{\arrjoin{a_2:n_2}{\varphi}}
}

\texttt{arrayJoin} is commutative over distinct attributes: changing the order does not affect the result.

\subsubsection{Pushing Down Filter under ArrayJoin}
\ruleeq[.1]{
\filt{\arrjoin{a_i:n_i}{\varphi}} \eq \arrjoin{a_i:n_i}{\filt{\varphi}}
}
A filter $\sigma_\theta$ can be pushed below \texttt{arrayJoin} $\mu$ if $\theta$ does not reference the flattened columns $n_i$ introduced by the \texttt{arrayJoin}.  
This is a \textbf{rule-based} transformation as filter pushdown reduces intermediate data, whereas \texttt{arrayJoin} multiplies rows by array lengths.
\subruleeq[.2]{
\filt{\arrjoin{a_i:n_i}{\varphi}} \eq 
\arrjoin{a_i:n_i}{\arrfilts[(a_i:n_i,\theta)]{\varphi}}
}
When a filter $\sigma_\theta$ targets columns produced by \texttt{arrayJoin}, it can be rewritten as an \texttt{arrayFilter} on the original arrays.  
To preserve element-wise correspondence, the \texttt{arrayFilter} is applied jointly across all columns involved.  
This \textbf{rule-based} transformation reduces the number of elements flattened, minimizing intermediate results, particularly for wide arrays.

\subsubsection{Eliminating Empty Arrays Before arrayJoin}
\subruleeq[.3]{
\arrjoin{a:n}{\varphi} \eq \arrjoin{a:n}{\filt[a!= \lbrack \rbrack]{\varphi}}
}
This rule filters out empty arrays before applying \texttt{arrayJoin} operator, reducing unnecessary expansions.
This optimization is \textbf{cost-based}, as its effectiveness depends on the selectivity of the filter—if most arrays are empty, the gain is significant; otherwise, the overhead of filtering may outweigh the benefit.

\subsubsection{Pushing Down Projection under ArrayJoin}
\ruleeq{
\proj{L \cup \{n_i\}}{\arrjoin{a_i:n_i}{\varphi}} \eq \proj{L \cup \{n_i\}}{\arrjoin{a_i:n_i}{\proj{L \cup \{a_i\}}{\varphi}}}
}
When a projection selects all columns produced by \texttt{arrayJoin}, it can be pushed below the operator, replacing transposed columns $n_i$ with the original arrays $a_i$.  
It is a \textbf{rule-based} transformation as it reduces the number of columns that have to be materialized by the arrayJoin operator.

\subsubsection{Pushing Down Join under ArrayJoin}
\ruleeq[.1]{
\arrjoin{a_i:n_i}{\varphi_1} \join \varphi_2  \eq \arrjoin{a_i:n_i}{\varphi_1 \join \varphi_2}
}
%
The join with $\varphi_2$ can be performed either before or after the \texttt{arrayJoin} on $\varphi_1$, depending on its effect on intermediate result size.  
This is a \textbf{cost-based} decision: if the join reduces or preserves the size of its inputs, it can be applied first; otherwise, the choice depends on the relative selectivity of the join compared to the number of elements in $a_i$.  
\subruleeq[.2]{
\arrjoin{a_i:n_i, b_j:m_j}{\varphi_1 \join \varphi_2} \eq
\begin{aligned}[t]
& \arrjoin{a_i:n_i,I}{\derive[I=arrayEnumerate(a_1)]{\varphi_1}} \\
& \join \arrjoin{b_j:m_j,I}{\derive[I=arrayEnumerate(b_1)]{\varphi_2}}
\end{aligned}
}

For an \texttt{arrayJoin} applied to corresponding attributes $a_i$ and $b_j$ in $\varphi_1$ and $\varphi_2$, the transformation uses \texttt{arrayEnumerate} to generate indices $[1,\dots,\texttt{length}(a_i)]$ and enforces their equality, ensuring aligned matching of array elements.

\subsubsection{Pushing Down Derive under ArrayJoin}


\ruleeq[.1]{
\derive{\arrjoin{a_i:n_i}{\varphi}} \eq \arrjoin{a_i:n_i}{\derive{\varphi}}
}

When the derived expression does not involve the array elements introduced by the \texttt{arrayJoin}, the \texttt{derive} operation can be pushed below the join without any modification. This transformation is \textbf{rule-based}, as it avoids redundant computation over the expanded rows generated by the \texttt{arrayJoin}.
\subruleeq[.2]{
\derive[y=f(n)]{\arrjoin{a:n}{\varphi}} \eq \arrjoin{y}{\derive[y=\texttt{arrayMap}_{(f)}(a)]{\varphi}}
}

In the case where the derived column $y$ depends on the array elements produced by the \texttt{arrayJoin}, the transformation remains valid by rewriting the expression using $\texttt{arrayMap}_{(f)}$.

\subsubsection{Pushing down ArrayFilter under ArrayJoin}

\ruleeq{
\arrfilt[b:m,\theta]{\arrjoin{a_i:n_i}{\varphi}} \eq \arrjoin{a_i:n_i}{\arrfilt[b:m,\theta]{\varphi}}
}
An \texttt{arrayFilter} can be pushed below \texttt{arrayJoin} when the filtered array $b$ is independent of the arrays $a_i$ being flattened.  
This \textbf{rule-based} transformation is more efficient, since filtering before expansion avoids generating unnecessary rows.

\subsubsection{Commutativity of ArrayFilter}

\ruleeq[]{
{\arrfilt[a_2:n_2,\theta_2]{\arrfilt[a_1:n_1,\theta_1]{\varphi}} \eq 
\arrfilt[a_1:n_1,\theta_1]{\arrfilt[a_2:n_2,\theta_2]{\varphi}}}
}
\texttt{arrayFilter} operators commute: filtering on $a_1$ and $a_2$ yields the same result regardless of order.  
Each filter acts independently on its array, without interfering with others.





\subsubsection{Pushing Down Filter under ArrayFilter}

\ruleeq[]{
 \filt{\arrfilt[a:n,\theta_a]{\varphi}} \eq \arrfilt[a:n,\theta_a]{\filt{\varphi}}
}
A global filter $\sigma_\theta$ can be pushed below an \texttt{arrayFilter} when $\theta$ is independent of the array condition $\theta_a$.  
This \textbf{rule-based} transformation is always beneficial: the filter reduces input rows before array processing, while \texttt{arrayFilter} only shrinks arrays horizontally.



\subsubsection{Pushing Down Projection under ArrayFilter}

\ruleeq{
\proj{L \cup \{n\}}{\arrfilt{\varphi}} \eq 
\proj{L \cup \{n\}}{\arrfilt{\proj{L \cup \{a\}}{\varphi}}}
}
A projection can be pushed below \texttt{arrayFilter} if it preserves the source array $a$ used to derive $n$.  
This \textbf{rule-based} transformation is always safe and efficient: projecting early reduces columns processed and minimizes data access.

\subsubsection{Distribution of ArrayFilter over Join}
\ruleeq[.1]{
  \arrfilt[(a:n,\theta)]{\varphi_1 \join \varphi_2} \eq \arrfilt[(a:n,\theta)]{\varphi_1} \join \varphi_2 
}
The arrayFilter operation distributes over a join when the filtered attribute $a$ involves only attributes from one of the joined expressions, or for the case of corresponding arrays $a_i$ as shown in \ref{R10.2}:
\subruleeq[.2]{
\arrfilts[(a_1:n_1,\dots,a_k:n_k,\theta)]{\varphi_1 \join \varphi_2} \eq \arrfilts[(a_1:n_1,\dots,a_k:n_k,\theta)]{\varphi_1} \join \varphi_2
}
For \texttt{arrayFilter} applied on two attributes \(a\) and \(b\) involving \(\varphi_1\) and \(\varphi_2\), respectively, \texttt{arrayFilter} can be distributed over the join by separating the \texttt{arrayFilter} across the two expressions, provided there is no correspondence between the two columns \(a\) and \(b\). 
\subruleeq[.3]{
\arrfilts[(a:n,\theta_{a}),(b:m,\theta_{b})]{\varphi_1 \join \varphi_2} 
\eq 
\arrfilt[(a:n,\theta_{a})]{\varphi_1} \join \arrfilt[(b:m,\theta_{b})]{\varphi_2}
}
These are a \textbf{cost-based} transformations: the decision to push down the arrayFilter depends on the selectivity and complexity of the filter condition, as well as the selectivity of the join. 

\subsubsection{Commutativity of  ArrayFilter and Derive}

\ruleeq[.1]{
\arrfilt{\derive{\varphi}}  \eq \derive{\arrfilt{\varphi}}
}
\noindent
This equivalence holds when the derived column $y = f(x_1, \dots, x_s)$ is computed independently of the array $a$ being filtered by the \texttt{arrayFilter} condition $\theta$. 

\vspace{-1em}
\subruleeq[.2]{
\arrfilt[y:n,\theta]{\derive[y=\texttt{arrayMap}_{(f)}(a)]{\varphi}}  \eq \derive[y=\texttt{arrayMap}_{(f)}(a)]{\arrfilt[a:n,\theta^{'}]{\varphi}}
}
When an \texttt{arrayFilter} is applied to a derived array $y$ and the predicate is invertible ($\theta(y) \equiv \theta'(a)$, see Definition~\ref{invertible-filter}), it can be pushed below the derive.  
This \textbf{rule-based} transformation avoids computing $f$ on irrelevant elements; if $y$ is unused later, the derive may also be eliminated.



\subsubsection{Commutativity of Derive}

\ruleeq[]{
  \scalebox{0.92}{
$\derive[y_2=f_2(z_1,\dots,z_j)]{\derive[y_1=f_1(x_1,\dots,x_i)]{\varphi}} \eq 
\derive[y_1=f_1(x_1,\dots,x_i)]{\derive[y_2=f_2(z_1,\dots,z_j)]{\varphi}}$}
}
The derivation operation is commutative when the columns on which the second derivation is applied do not depend on the columns resulting from the first derivation, i.e., $y1 \notin \{z_1,\dots,z_j\}$.




\subsubsection{Pushing Down Filter under Derive}

\ruleeq[.1]{
\filt{\derive{\varphi}} \eq \derive{\filt{\varphi}}
}
When the filter $\theta$ does not reference the derived column $y$, it can be pushed below the \texttt{derive} operation.  
This \textbf{rule-based} transformation reduces the number of rows before derivation.

\subsubsection{Invertibility of Filter Applied to Derived Column}

\subruleeq[.2]{
{\filt[\theta(y)]{\derive[y=f(x)]{\varphi}} \eq \derive[y=f(x)]{\filt[\theta^{'}(x)]{\varphi}}}
}
When a filter is applied to a derived column $y = f(x)$, and the predicate $\theta(y) = \theta \circ f(x)$ is invertible (Definition \ref{invertible-filter}), the filter can be rewritten and pushed below the derive operator by applying the inverse transformation directly on the original column $x$.

This \textbf{rule-based} transformation pushes the filter down to the original column thereby reducing intermediate data and enabling the execution engine to leverage existing indexes.
Furthermore, if the derived column \(y\) is used solely for filtering and is not projected later, it can be eliminated, avoiding unnecessary computations.

\subsubsection{Pushing Down Projection under Derive}

\ruleeq{
\scalebox{0.92}{$\proj{L \cup \{y\}}{\derive{\varphi}} \eq \proj{L \cup \{y\}}{\derive{\proj{L \cup \{x_1,\dots,x_s\}}{\varphi}}}$}
}
A projection on a column $y=f(x_1,\dots,x_s)$ can be pushed below \texttt{derive} if it retains all columns $x_1,\dots,x_s$.  
This \textbf{rule-based} transformation reduces the data processed in subsequent operations.

\subsubsection{Distribution of Derive over Join}

\ruleeq[]{
\derive{\varphi_1 \join
\varphi_2} \eq \derive{\varphi_1} \join \varphi_2
}
The \texttt{derive} operation can be pushed below a join when all its input columns $x_1,\dots,x_s$ come from a single input (e.g., $\varphi_1$).  
This is a \textbf{cost-based} transformation: if the join significantly increases row count (e.g., many-to-many), pushing down \texttt{derive} reduces data processed; otherwise, applying the join first is more efficient.

\subsection{Aggregation rewrite rules}

\subsubsection{Swapping of filtering with aggregation:}

\ruleeq{
{\filt[\theta_G]{\agg{\varphi}} \eq \agg{\filt[\theta_G]{\varphi}}}
}

The filter can be pushed down under the \texttt{aggregation} if the columns being filtered by $\theta_G$ are a subset of the grouping columns $G$. This transformation is \textbf{cost-based}, as its benefit depends on the selectivity of the filter and the granularity of the aggregation.

\begin{defi}[Distributive aggregation]\label{def:distagg}
  An Aggregation function $\aggf$ is said to be distributive when there exist a function $F_{\aggf}$, a function $T_{\aggf}$, and an associative operator $\oplus_\texttt{\aggf}$ such that
     $$\aggf{\bag{e_1, ..., e_n}} = T_{\aggf}(F_{\aggf}(e_1) \oplus_\texttt{\aggf} ... \oplus_\texttt{\aggf} F_{\aggf}(e_n))$$

\end{defi}

\begin{example}
  $\texttt{sum}$, $\texttt{Avg}$, and $\texttt{distinct}$ are distributive aggregations, and we have: 
  \begin{itemize}
  \item $F_{\texttt{sum}}$ and $T_{\texttt{sum}}$ are the identity function and $\oplus_{\texttt{sum}} = +$.

  \item $F_{\texttt{Avg}}: e \rightarrow (e, 1)$, $T_{\texttt{Avg}}: (s,c) \rightarrow s/c$, and $ (s_1, c_1) \oplus_{\texttt{Avg}} (s_2, c_2) = (s_1 + s_2, c_1 + c_2)$
  
  \item  $F_{\texttt{distinct}}: e \rightarrow \bag{e}$, $T_{\texttt{distinct}} = id$ , and $\oplus_{\texttt{distinct}}~=~\cup$
  \end{itemize}
\end{example}

Note that any distributive aggregation $\aggf$ can be decomposed into two aggregations denoted $\aggfirst$ and $\aggfin$, where $\aggfirst = \texttt{reduce}(\oplus, \texttt{map}(F_{\aggf},.))$ and $\aggfin = T_{\aggf}(\texttt{reduce}(\oplus, .))$. Table~\ref{tab:decomposition} illustrates this decomposition for common aggregation functions.
We suppose in the following that all aggregations are distributive.

\begin{footnotesize}
  \begin{table}
    \centering

      \begin{tabular}{c|c|c}  
          \textbf{Agg} & \textbf{$\text{Agg}^{(i)}$} & \textbf{$\text{Agg}^{(f)}$} \\ 
          \hline
          min & min & min \\ 
          max & max & max \\ 
          count & count & sum \\ 
          sum & sum & sum \\ 
          avg & sum, count & sum, sum \\ 
      \end{tabular}
      \caption{Decomposition of aggregate functions}
    \label{tab:decomposition}
       \vsqueezeaftercaption
       \vsqueezeaftercaption
\end{table}
\end{footnotesize}

\subsubsection{Introducing Pre-aggregation under ArrayJoin}

\begin{enumerate}
  \item Aggregating flattened array column by a scalar column: using $\aggforeach{}$ as a pre-vertical aggregation that aggregates the corresponding array items position-wise, then applying \texttt{arrayAgg()} function that aggregate array elements: 
\ruleeq[.1]{
 \scalebox{0.95}{$
 \agg[s,agg(a):n]{\arrjoin{A:a}{\varphi}} \eq 
 \derive[n=arrayAgg(N)]{\agg[s,aggForEach(A):N]{\varphi}}
 $}
}

\item Aggregating scalar column by flattened array column: perform \texttt{pre-agg} operation before flattening the grouping array column:
\subruleeq[.2]{
 \scalebox{1}{$
 \agg[a,agg(s):n]{\arrjoin{A:a}{\varphi}} \eq 
 \agg[a,agg(s):n]{\arrjoin{A:a}{\agg[A,agg(s):n]{\varphi}}}
 $}
}

\item Aggregating flattened array column by its corresponding flattened array column: using $\aggforeach{}$ as a pre-vertical aggregation that aggregates the corresponding array items position-wise, then applying the final aggregation operation after flattened the corresponding arrays: 
\subruleeq[.3]{
 \begin{aligned}
 &\agg[a_1,agg(a_2):n]{\arrjoin{A_1:a_1,A_2:a_2}{\varphi}} \eq \\
 &\agg[a_1,agg(n):n]{\arrjoin{A_1:a_1,N:n}{\agg[A_1,aggForEach(A_2):N]{\varphi}}}
 \end{aligned}
}
\end{enumerate}


\subsubsection{Introducing Pre-aggregation under Filter:}

\ruleeq[]{
  \begin{aligned}
&\agg[G, agg(X):m]{\filt[\theta_{L}]{\varphi}} \eq \\
&\agg[G, \aggfin{m}:m]{\filt[\theta_{L}]{\agg[G \union L, \aggfirst{X}:m]{\varphi}}}
  \end{aligned}
}

 This rule introduces a pre-aggregation under the \texttt{filter} operator by adding the filtered columns $L$ to the grouping columns of the inner aggregation. This rule is applicable
   when the filter columns do not intersect with the aggregated metrics $X$.

\subsubsection{Introducing Pre-aggregation under Derive:}

\ruleeq[]{
  \begin{aligned}
&\agg[G, agg(X):m]{\derive{\varphi}} \eq \\
&\agg[G, \aggfin{m}:m]{\derive{\agg[G \union \{x_1, \dots, x_s\}, \aggfirst{X}:m]{\varphi}}}
  \end{aligned}
}

This rule introduces a pre-aggregation under the \texttt{derive} operator by adding the input columns $(x_1,\dots,x_s)$ of the derivation function $f$ to the grouping columns of the inner aggregation. If $f$ is injective, the final aggregation can be omitted.

\subsubsection{Introducing Pre-aggregation Under arrayFilter}

  \ruleeq[]{
   \begin{aligned}
  &\agg[G,agg(X):m]{\arrfilt{\varphi}} \eq\\
 &\agg[G,\aggfin{m}:m]{\arrfilt{ \agg[G\union \{a\},\aggfirst{X}:m]{\varphi}}}
     \end{aligned}
  }

This rule introduces a pre-aggregation under the \texttt{arrayFilter} operator by adding the column $a$ to the grouping columns of the inner aggregation. 

\subsubsection{Distribution of aggregation over join} 
\ruleeq[]{
 \begin{aligned}
  &\agg[G_1 \union G_2, \aggf{X}:n]{\varphi_1 \join \varphi_2} \eq \\
  &\agg[G_1 \union G_2, \aggfin{m}:n]{\agg[G_1 \union \{jk\},\aggfirst{X}:m]{\varphi_1} \join \varphi_2}
 \end{aligned}
}

This rule introduces a pre-aggregation below the \texttt{join} when the aggregated attributes \(X\) belong exclusively to one join operand. 
It decomposes the aggregation into two stages: a local pre-aggregation (\(\aggfirst\)) on \(\varphi_1\) grouped by \(G_1 \cup \{jk\}\), followed by a final aggregation (\(\aggfin\)) after the join.

The previous transformations (R17.1-R21) are \textbf{cost-based}. Aggregations significantly reduce
intermediate results when the granularity is much smaller than  input size ($|G| \ll |\varphi|$, which is most often the case in practice with large datasets).  Otherwise, the benefit of these transformations depend on the aggregation granularity, the complexity of the derive function, and the filter's selectivity.

\section{Exploration of Query Plans}
Equivalence-preserving transformation rules, such as those proposed in Sec.~\ref{sec:rules}, can always be implemented within a Volcano-style optimization framework~\cite{graefe-tkde94}. Volcano’s transformational engine systematically explores alternative query plans by applying rewrite rules until no new expressions are generated. While this approach is general and extensible, the number of possible rewritings grows exponentially with query size, making exhaustive enumeration impractical for complex queries.
Another option is to use a greedy optimizer which uses the rule categorisation presented in Sec.~\ref{sec:rules}. Whenever a transformation rule pattern is detected, it systematically applies it if it is rule-based, and if it is cost-based it tests whether the cost condition is satisfied in order to apply it. 
Such an optimizer would be much faster but lacks an overall optimality guarantee as it uses local decisions only.

As a third strategy, we propose an enumeration method that provides an optimality guarantee, and whose complexity is polynomial to the number operators other than joins. %

\subsection{Overall Optimization Process}\label{sec:algooverview}

The optimization process proceeds in three main stages: pre-processing, enumeration, and post-processing. 

\subsubsection{Pre-processing}
The purpose of preprocessing is to prepare the query plan for the subsequent enumeration stage, in particular by applying transformations that introduce new operators, so that the subsequent enumeration stage only needs to consider operator reordering.
Projections are first pulled to the top of the query tree. Next, all transformations which introduce new operations are applied (Rules~\ref{R2.2}, \ref{R2.3}).
To explore all opportunities to apply \ref{R2.2}, we proceed as follows. For each array column $a$, we push all filters $\filtN[\theta_a]$ downward as much as possible and, if necessary, pull $\arrjoinN$ upward until a subterm of the form $\filt{\arrjoin{a}{\varphi}}$ is reached. If such a term is encountered, the rule is applied; otherwise it is not applicable.
Applying \ref{R2.3} simply consists in inserting a filter that removes empty arrays below \texttt{arrayJoin} operators.

 \subsubsection{Enumeration}\label{sec:enumeration}
 The enumeration stage aims to compute an execution plan with an optimal ordering of operators. It builds on earlier work on optimizing join queries in the presence of expensive filters~\cite{chaudhury99}. 
 The core idea is to replace exhaustive exploration of filter permutations with a ranking-based order inspired by results from the task scheduling domain~\cite{monma81,smith1956}.
We adapt this approach to the A3D-RA algebra by (i) generalizing ranking to unary operators other than filters, such as derive, array filters, and array joins, and (ii) accounting for precedence constraints between operators, such as: a derive operation which produces column $c$ needs to take place before any operation which uses $c$. We then combine this ranking of unary operators with a state-of-the-art top-down join enumeration technique \cite{dehaan2007,fender2012} which we adapt to take derive operations (which can be used to compute new join keys) into account. 
This design promotes extensibility: new unary operators can be integrated by defining ranking and precedence constraints, while different join enumeration strategies can be plugged in independently.

\subsubsection{Post-processing}
Finally, the post-processing step applies pre-aggregation rules--i.e. aggregation rules that introduce pre-aggregations (see Sec.~\ref{sec:agg})-- to the optimal term identified during enumeration, producing the final optimized query plan.
Rule application proceeds iteratively from the selected plan until no further aggregation rule is applicable. Pre-aggregation rules are deferred to this phase for two reasons. First, they introduce new operators and are therefore not considered during the enumeration phase, which is dedicated to operator reorderings. Second, pre-aggregations can be blocking and thus interfere with the exploration of reorderings. Deferring their application ensures that the enumeration phase can consider all relevant reorderings of non-blocking operators.
In practice, this design choice is further justified by the fact that aggregation operators are typically few in number and often appear near the root of the query plan, making a dedicated post-processing phase both effective and inexpensive.

\subsection{Enumeration Method}

We now describe the enumeration stage in more details.
\subsubsection{Enumeration Algorithm}

Enumeration is performed by Algorithm~\ref{alg:enumerate}, which takes as input the initial query $Q$ and a join graph $G$ where a node represents a relation and an edge links two nodes when a join predicate exists between them. Joins are enumerated top-down by recursively partitioning the graph (\texttt{Partition} method is detailed in Sec.~\ref{sec:joinenum}). 

When enumerating a join, the original~\cite{chaudhury99} optimizer identifies the filters that are applicable beneath that join, sorts them by rank, and generates plans where only the first $i$ filters are pushed below the join, for all $i$. Algorithm~\ref{alg:enumerate} extends this approach by considering all unary operators applicable below each side of the join (lines 13-14). These operators are sorted according to the ranking strategy described in Sec~\ref{sec:sortops}. Since we have derive operators that can generate join keys, these operators need to be performed before the joins using those keys. This is why the algorithm (in lines 15 and 16) computes the index $oi$ of  the last operator that needs to be performed before the join, then computes plans where the first $i$ operators are pushed with $i$ starting from $oi$. 
The presence of \texttt{derive} operations can also lead to infeasible joins, i.e., joins whose key is produced by different relations that do not belong to the same subset ($p_i$). All such cases are detected by the \texttt{Valid} function (line 9), which checks for the existence of an operator that must be applied before the join but cannot be applied to either side of it.

\textit{best} is a memoization structure used to store the optimal plan corresponding to each enumerated combination of (1) set of joined base relations and (2) set of operators applied within the associated join tree. It is used to construct bigger expressions using already computed sub-expressions and thus avoids redundant computations.
$\textit{best}[p, t]$ stores the optimal plan that joins the relations in~$p$ while applying the operators in~$t$ under the root join, whereas $\textit{best}[p]$ denotes the collection of $\textit{best}[p, t]$ for all enumerated $t$.

\subsubsection{Sorting unary operations}\label{sec:sortops}

 \begin{algorithm} 
\begin{small}
\caption{Enumeration of Algebraic Terms}
\label{alg:enumerate}
\DontPrintSemicolon

\SetKwFunction{Enumerate}{Enumerate}
\SetKwFunction{EnumerateAgg}{EnumerateAgg}
\SetKwFunction{Valid}{Valid}
\SetKwFunction{Continue}{Continue}
\SetKwFunction{Applicable}{Applicable}
\SetKwFunction{Cost}{cost}
\SetKwFunction{Partition}{Partition}

\Enumerate{$G$, $Q$}: \;
\If{$best[G] \neq \emptyset$}{ 
    \Return \tcp*[f]{already been computed}
}

\If{$G$ contains a single node}{ 
    $best[G] \gets G$

    \Return
}
\ForEach{\Partition $(p_1, p_2)$ of $G$}{
    \Enumerate{$p_1$, $Q$}; \Enumerate{$p_2$, $Q$}\;
    \If{$\lnot$ \Valid{$p_1, p_2$}}{
        \Continue
    }
    \ForEach{$s \in best[p_1]$}{
        \ForEach{$t \in best[p_2]$}{
            $o_1 \gets$ \Applicable{$s$, $Q$} \tcp*[f]{sorted applicable operators on $s$}
            
            $o_2 \gets$ \Applicable{$t$, $Q$}
            
            $oi_1 \gets$ last index of operator that must be applied on $s$\;
            
            $oi_2 \gets$ last index of operator that must be applied on $t$\;

            \For{$i \gets oi_1$ \KwTo $|o_1|$}{
                \For{$j \gets oi_2$ \KwTo $|o_2|$}{
                    $join \gets (o_1[0..i](s)) \Join (o_2[0..j](t))$\;
                    
                    \If{$\Cost{join} < best[G, ops(join)]$}{
                        $best[G, ops(join)] \gets join$\;
                    }
                }
            }
        }
    }
}
\end{small}
\end{algorithm}

Consider a sequence of filters,
$\filt[\theta_1]{\filt[\theta_2]{\ldots\filt[\theta_n]{R}\ldots}}$.
Assume that each $\theta_i$ has an average computing cost per
tuple $c_i$ and average selectivity $s_i\in[0,1]$, such that for any relation $S$, we can estimate that $\filt[\theta_i]{S}$ costs $c_i|S|$ to compute and yields a result containing $s_i|S|$ tuples. Then it has been proved~\cite{hellerstein93,chaudhury99} that, when the filters are independent, the overall cost of the sequence is minimized by applying them in descending order of their \emph{ranks} $\frac{1-s_i}{c_i}$.

 Applying this ranking metric to \ra{} operators would assign rank $0$ to all 
 operators which do not reduce the number of tuples in the input relation, since their selectivity is $1$. Consequently, they would all be applied last, in arbitrary order. However, we can do better by noticing that, while operators such as array filters do not reduce the number of tuples, they reduce the size of the array columns on which they operate.
 The cost of an operator therefore depends not only on the vertical selectivity of preceding operators, but also on what we call their \emph{horizontal selectivity}—that is, how much they reduce the size of the arrays. 
Using these metrics, we define the following relation $\lesssim$ on unary operators:
\begin{equation*}
    i \lesssim j \iff 
    \begin{cases}
    i \in \{\filtN[], \aggN[]\}, j \in \{\arrfiltN[], \deriveN[]\}, i \text{ and }j \text{ process different arrays}\\
     r_i \geq r_j, \text{ where }
    r_i = \begin{cases}
    \frac{1-s_i}{c_i} \text{ when } i\in{\filtN[], \aggN[]}\\
    \frac{1-s_i^a}{c_i} \text{ when } i\in{\arrfiltN[\theta_a], \deriveN[y=f(a)]}\\
    \frac{1-|a|}{c_i} \text{ when } i= \arrjoinN
    \end{cases}
    \end{cases}
\end{equation*}
where $s_i$ denotes the vertical selectivity of operator $i$, $c_i$ its cost per tuple (or per array element when operating on an array column), $s_i^a$ its horizontal selectivity on array $a$, and $|a|$ the average size of array column $a$.
Selectivity and per-tuple cost estimation is detailed in Sec.~\ref{sec:cost-model}.

 Ordering operators by $\lesssim$ yields an optimal order under the assumptions about cost and selectivity which we detail in section~\ref{sec:optimality}.
However, a simple sorting is not always applicable because unary operators cannot be applied in just any order. For instance, a filter on column $c$ cannot precede a \texttt{derive} operator that produces $c$. To handle such dependencies, we follow techniques from the task scheduling literature~\cite{monma81}.

\paragraph{Precedence constraints.}
Semantic restrictions constraining the order of operators are naturally expressed as \emph{precedence constraints}, forming a directed acyclic graph (DAG) called a \emph{precedence graph}. An edge from an operator $o_1$ to an operator $o_2$ means that $o_1$ needs to be applied before $o_2$. When the precedence graph is \emph{series-parallel}~\cite{monma81}, efficient polynomial-time algorithms exist for constructing an order that is consistent with precedence constraints while minimizing cost. Beyond series-parallel constraints, the problem is shown to be NP-hard~\cite{kelly82}.
A precedence graph is not series-parallel if and only if it contains four nodes in a \emph{Z}-shaped relation~\cite{DROR20101767,lawler2006}, i.~e. with precedence constraints $A < C$, $B <C$ and $B < D$. In such a structure, the optimal solution may have $A$ before $B$ or the reverse. Adding an edge between them in either direction eliminates the $Z$ and restores the series-parallel property, but potentially at the cost of optimality. In our solution, we consider the following heuristic: when a Z-structure is detected, we add an edge from $o_1$ to $o_2$ if $o_1 \lesssim o_2$, and an edge from $o_2$ to $o_1$ otherwise.

\null

\subsubsection{Join enumeration}~\label{sec:joinenum}
We use join graphs for cross-product-free top-down join enumeration~\cite{dehaan2007,fender2012}. These works propose efficient graph partitioning methods for dividing the join graph (\texttt{partition} function of algorithm~\ref{alg:enumerate}) into two subsets. 
In classical relational algebra, the join graph is a graph where a node represents a base relations, and an edge represents a join predicate between two nodes. In our extended algebra, there are cases where a join predicate involves a column produced by a derive operator which might use columns coming from more than one base relation. 
To make previous techniques applicable to our algebra, we extend the join graph construction in the following way: for each join predicate, we determine the sets of relations $L$ and $R$ that are involved in the left and right side of the join respectively. An edge $(l,r)$ is then produced for all $l \in L$ and $r \in R$. 
This way, no join is missed. However, unfeasible joins can be produced and are detected using the \texttt{Valid} function mentioned in Sec.~\ref{sec:algooverview}.

\subsubsection{Rank Estimation}
\label{sec:cost-model}
Enumeration relies on a per-operator rank determined by its selectivity and per-tuple cost. 
The cost $c$ of a unary operator is estimated as $c = c_t \times |R|$,
where $c_t$ denotes the operator’s per-tuple cost and $|R|$ the cardinality of its input relation.
In the context of $\ra{}$, the dominant component of the per-tuple cost typically arises from iterating over array elements. Accordingly, we approximate $c_t$ by the average length of the array column processed by the operator, and by~1 when the column is atomic.
Refined estimations of per-tuple costs is beyond the scope of this paper; such estimates can also be provided by the user.

Selectivity estimation relies on the notion of the \emph{relative frequency} of a value $v$ in a column $C$, defined as the ratio between the number of occurrences of $v$ and the total number of rows in $C$.
To improve the accuracy of selectivity estimates, we account for the underlying data distribution of each column. A dedicated statistics module maintains different statistics depending on the data distribution:  for low-cardinality columns, exact statistics are maintained in the form of a mapping $(v \mapsto fr)$.  For columns whose values are nearly uniformly distributed, only the average frequency is stored.  For skewed distributions (normal, left-skewed, right-skewed), clustering is applied using the K-Means algorithm to group values into clusters of low intra-dispersion. 
\halonly{Data-clustering algorithm is given in Appendix~\ref{appendix:costmodel}.}
  For array-typed columns,  two types of statistics are captured: 
         \emph{Array statistics} ($stats^{array}$), representing the array distribution, and 
         \emph{Row statistics} ($stats^{row}$), representing the row distribution.
Using these statistics, selectivities are estimated for different types of predicates, and for atomic and array columns.
\halonly{Detailed selectivity formulas are given in Appendix~\ref{appendix:costmodel}.}

\subsubsection{Optimality}\label{sec:optimality}
We show that the ordering computed by Algorithm~\ref{alg:enumerate} is optimal. %
 The idea of ordering filters by rank~\cite{hellerstein93} originates in the task scheduling literature~\cite{smith1956,monma81}, where Smith’s theorem establishes optimal scheduling for cost functions satisfying the \emph{Adjacent Pairwise Interchange (API)} property.
\begin{ppty}[API property]\label{prop:api}
A cost function $f$ satisfies the API property if there exists a transitive and complete binary relation $\lesssim$ such that, for any jobs $i$ and $j$,
\[
i \lesssim j \implies f(u,i,j,v) \leq f(u,j,i,v) \quad \text{for all sequences } u,v.
\]
Here, $f(u,i,j,v)$ denotes the cost of executing sequence $u$, followed by $i$, then $j$, then $v$.
\end{ppty}

Smith’s theorem states that if $f$ satisfies the API property, any permutation consistent with $\lesssim$ is \textbf{optimal}.
Assume the following:
\begin{itemize}
\item the columns involved in filtering predicates are statistically independent;
\item when a filter is applied on a column that is used as an aggregation key, it does not affect the average number of tuples per value of the key;
\item the cost of aggregation is proportional to the input size (as is typically the case for hash-based aggregation algorithms). 
\end{itemize}
Under these assumptions, we show that the API property holds for the relation $\lesssim$ (Sec.~\ref{sec:sortops}) together with the cost model of the unary operators $\filtN[]$, $\arrfiltN[]$, $\arrjoinN[]$, $\deriveN[]$, and $\aggN[]$ which we call \emph{rankable operators}. %
\begin{proof}[Proof sketch]
    Let $i$ and $j$ be two operators such that $i \lesssim j$, and let $u$ and $v$ be any two sequences of operators. We recall that $s_o$ denotes the selectivity of operator $o$ and $c_o$ its cost per tuple.

    When $i$, and $j$ are filters, =
    we have $f(u,i,j,v) =  f(u) + (\prod_{o \in u}s_o)c_i|R| + (\prod_{o \in u}s_o)s_i c_j|R| + f_{u,i,j}(v)$ and $f(u,j,i,v) = f(u) + (\prod_{o \in u}s_o)c_j|R| + (\prod_{o \in u}s_o)s_j c_i|R| + f_{u,i,j}(v)$. $f_{u,i,j}(v)$ denotes the cost of $v$ after applying the sequence $u,i,j$, and $|R|$ denotes the size of the input relation. So $f(u,i,j,v) - f(u,j,i,v) = c_i(1-s_j)|R| - c_j(1-s_i)|R| \leq 0$ because $\frac{1-s_j}{c_j}\leq\frac{1-s_i}{c_i}$.
\\
    When $i$ is a filter and $j$ is an array-filter that process different arrays we have $f(u,i,j,v) =  f(u) + (\prod_{o \in u}s_o)c_i|R| + (\prod_{o \in u}s_o)s_i c_j|R| + f_{u,i,j}(v)$, and $f(u,j,i,v) =  f(u) + (\prod_{o \in u}s_o)c_j|R| + (\prod_{o \in u}s_o)c_i|R| + f_{u,i,j}(v)$. So $f(u,i,j,v) - f(u,j,i,v) = c_j(s_i-1)|R| \leq 0$.
    Similarly, we can show the property for all combination of operations.
\end{proof}

Chaudhuri and Shim~\cite{chaudhury99} show that, to optimize a query containing both filters and joins, it suffices to consider all rewritings in which filters are ordered (joins may appear between them). The same argument extends to rankable operators beyond filters. Algorithm~\ref{alg:enumerate} enumerates exactly such rewritings.

In conclusion, under the assumptions stated above, Algorithm~\ref{alg:enumerate} finds an optimal ordering of A3D operators.

\subsubsection{Complexity}
\label{sec:complexity}

The original algorithm by Chaudhuri and Shim~\cite{chaudhury99} is proven to be polynomial with respect to the number of filters. The complexity regarding the number of joins depends on the chosen join enumeration strategy. Since join enumeration is a NP-hard problem, existing exhaustive join enumeration algorithms are exponential with respect to the number of joins. 
The adaptations we made to the original algorithm are all polynomial with respect to the number of operators. Hence, our approach is polynomial to the number of all operators except joins.

\usetikzlibrary{positioning}

\section{Experiments}
\label{sec:experiments}

We report on an experimental evaluation of A3D-RA. 

\subsection{System Architecture}
The A3D optimizer is designed as a modular, backend-independent, logical optimization layer that can be instantiated over different database systems. For a given backend system~$X$, we denote by A3DRA[$X$] the corresponding instantiation of our framework.

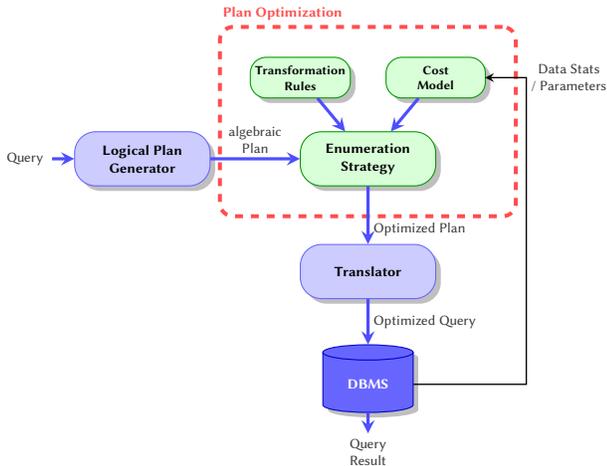
\begin{figure}[t]
    \centering
    \begin{tikzpicture}[scale=0.6, transform shape,
    component/.style={
        rectangle,
        rounded corners=8pt,
        minimum width=3cm,
        minimum height=1.2cm,
        align=center,
        draw=blue!60,
        fill=blue!20,
        font=\sffamily\bfseries,
        drop shadow
    },
    optimizer/.style={
        rectangle,
        rounded corners=8pt,
        minimum width=3cm,
        minimum height=1.2cm,
        align=center,
        draw=green!60!black,
        fill=green!15,
        font=\sffamily\bfseries,
        drop shadow
    },
    small_optimizer/.style={
        rectangle,
        rounded corners=6pt,
        minimum width=2.2cm,
        minimum height=0.9cm,
        align=center,
        draw=green!60!black,
        fill=green!15,
        font=\sffamily\small\bfseries,
        drop shadow
    },
    database/.style={
        cylinder,
        shape border rotate=90,
        aspect=0.3,
        minimum height=1.5cm,
        minimum width=2cm,
        align=center,
        draw=blue!80!black,
        fill=blue!60,
        text=white,
        font=\sffamily\bfseries,
        drop shadow
    },
    optim_box/.style={
        rectangle,
        dashed,
        draw=red!70,
        line width=1.5pt,
        rounded corners=5pt,
        inner sep=0.4cm
    },
    arrow/.style={
        ->,
        >=stealth,
        line width=1.2pt,
        draw=blue!70
    },
    thin_arrow/.style={
        ->,
        >=stealth,
        line width=0.5pt,
        draw=black
    },
    mylabel/.style={
        font=\sffamily\normalsize,
        text=black!80,
        align=center
    },
    redlabel/.style={
        font=\sffamily\bfseries,
        text=red!70
    }
]

\node[component] (logical) at (-1, 0) {Logical Plan\\Generator};
\node[optimizer] (enumeration) at (4, 0) {Enumeration\\Strategy};
\node[small_optimizer] (rules) at (2.5, 1.8) {Transformation\\Rules};
\node[small_optimizer] (cost) at (5.5, 1.8) {Cost\\Model};
\node[component] (translator) at (4, -2.5) {Translator};
\node[database] (dbms) at (4, -5) {DBMS};

\node[optim_box, fit={(rules) (cost) (enumeration)}] (optbox) {};
\node[redlabel, anchor=south west] at (optbox.north west) {Plan Optimization};

\node[mylabel, anchor=east] (query) at (-3, 0) {Query };
\node[mylabel] (result) at (4, -6.5) {Query \\ Result};

\draw[arrow] (query) -- (logical);
\draw[arrow] (logical) -- node[mylabel, above] {algebraic\\Plan} (enumeration);
\draw[arrow] (rules) -- (enumeration);
\draw[arrow] (cost) -- (enumeration);
\draw[arrow] (enumeration) -- node[mylabel, right, yshift=-9pt] {Optimized Plan} (translator);
\draw[arrow] (translator) -- node[mylabel, right] {Optimized Query} (dbms);
\draw[arrow] (dbms) -- (result);

\draw[thin_arrow] (dbms.east) -- ++(2.5,0) |- node[mylabel, pos=0.5, right] {Data Stats \\ / Parameters} ([xshift=2.2cm]cost.west);

\end{tikzpicture}
    \vsqueezeabovecaption\caption{A3D-Optimizer System Architecture.}
    \label{fig:archi}
    \vsqueezeaftercaption{}
\end{figure}

The architecture consists of three main components, as illustrated in Figure~\ref{fig:archi}.
The \emph{Logical Plan Generator} translates input queries into the A3D algebraic representation.
The \emph{Plan Optimization} module—comprising transformation rules, a cost model, and an enumeration strategy—explores the space of semantically equivalent plans to identify the estimated most efficient execution plan.
The cost model leverages data statistics and parameters provided by the target DBMS~$X$ to estimate the cost of candidate plans.
Finally, the \emph{Translator} converts the optimized algebraic plan back into an executable query in the native syntax of $X$.

\subsection{Experimental Protocol}
\label{sec:experimental-protocol}


\subsubsection{Considered dataset}
\label{sec:considered-dataset}

We consider a real-world use case from our industrial partner, which operates large-scale analytical workloads on financial data. The company maintains a denormalized, multidimensional schema designed to support interactive analytical queries involving grouping, filtering, and aggregation.


A distinctive feature of this industrial data model is the explicit use of \texttt{Array}-typed columns 
to represent repeated or hierarchical attributes—such as nested transaction codes, multi-level categorizations,  or composite financial indicators—within a single fact table. 
Each row may contain one or more arrays corresponding to additional analytical dimensions. 
This design eliminates costly joins while preserving the ability to perform complex multidimensional analyses directly within a columnar layout.

The considered dataset comprises over 100 million rows, exhibiting a high degree of heterogeneity across both scalar and array columns. 
Data distributions vary depending on business semantics and domain context, 
including \textit{uniform}, \textit{left-} and \textit{right-skewed}, and \textit{normal} patterns. 
Furthermore, array columns exhibit two distinct forms of variability: 
(i) array-level distribution, reflecting the diversity of distinct values per row, and 
(ii) row-level distribution, capturing the overall frequency of repeated values across rows.  
Array sizes also vary significantly across attributes. 
This multi-level variability, diversity and complexity of the dataset make it a challenging benchmark. 
It enables the evaluation of algebraic transformations across varying data distributions, allowing us to assess the real-world performance of the proposed framework in an authentic industrial setting.

\subsubsection{Real and Synthetic Query Workloads}
\label{sec:real-synthetic-workloads}

We evaluate A3D-RA on (i) 18 real-world analytical queries involving array flattening, array filtering, derived dimensions, and multidimensional aggregation, and (ii) a synthetic workload designed to isolate individual transformation rules under controlled selectivity and cardinality parameters. All queries are given at \cite{refExpDetails}. 

\subsubsection{Considered Backends}
\label{sec:baselines}

We instantiate A3D-RA over three state-of-the-art analytical database systems: ClickHouse~\cite{schulze2024clickhouse}, Umbra~\cite{neumann-umbra2020}, and Snowflake~\cite{snowflake2016}. 
These systems were selected because they consistently rank among the top-performing engines in recent large-scale analytical benchmarks~\cite{schulze2024clickhouse}. %
ClickHouse is a column-oriented analytical DBMS with native support for array-typed attributes. 
Umbra is a modern in-memory analytical system with an advanced cost-based optimizer and vectorized execution engine. 
Snowflake is a cloud-based analytical data warehouse supporting semi-structured data, including arrays.

\subsubsection{Comparative Analyses}
We first evaluate the quality of the generated execution plans, as measured by query runtime.
For each query and each considered backend system~$X$, we compare two configurations: $X$, where the original query is optimized and executed using the DBMS's built-in optimizer; and A3DRA[$X$], where the same query is first translated into A3D-RA, optimized using our algebraic framework, translated back into the native query language of~$X$, and then executed by the same DBMS engine.

Importantly, in both configurations the underlying execution engine remains unchanged. The only difference lies in the logical optimization phase. 
We then quantify the overhead introduced by A3D-RA by measuring its optimization time separately.

\subsubsection{Execution Setup}
\label{sec:execution-setup}
ClickHouse and Umbra experiments were conducted locally on a laptop equipped with an \texttt{AMD Ryzen\textsuperscript{TM}~5 PRO 3500U} processor with 
\texttt{24.0~GiB} of RAM, running \texttt{Ubuntu~25.04}.
Umbra was executed using the official Docker image (\begin{small}\texttt{umbradb/umbra:latest}\end{small}).
Snowflake experiments were conducted on the cloud using two warehouse configurations: \texttt{X-Small (XS)} (8~vCPUs, 16~GB RAM) and \texttt{Large (L)} (8×8~vCPUs, 8×16~GB RAM).
A timeout threshold of \texttt{1200}~s was set for each query execution. All reported execution times represent the average of three runs.

\subsection{Experimental Results}
\label{sec:experimental-results}
\subsubsection{Real-World Query Evaluation}
\label{sec:real-world-query-evaluation}

\begin{figure}[t]
    \centering
        \includegraphics[width=0.45\textwidth]{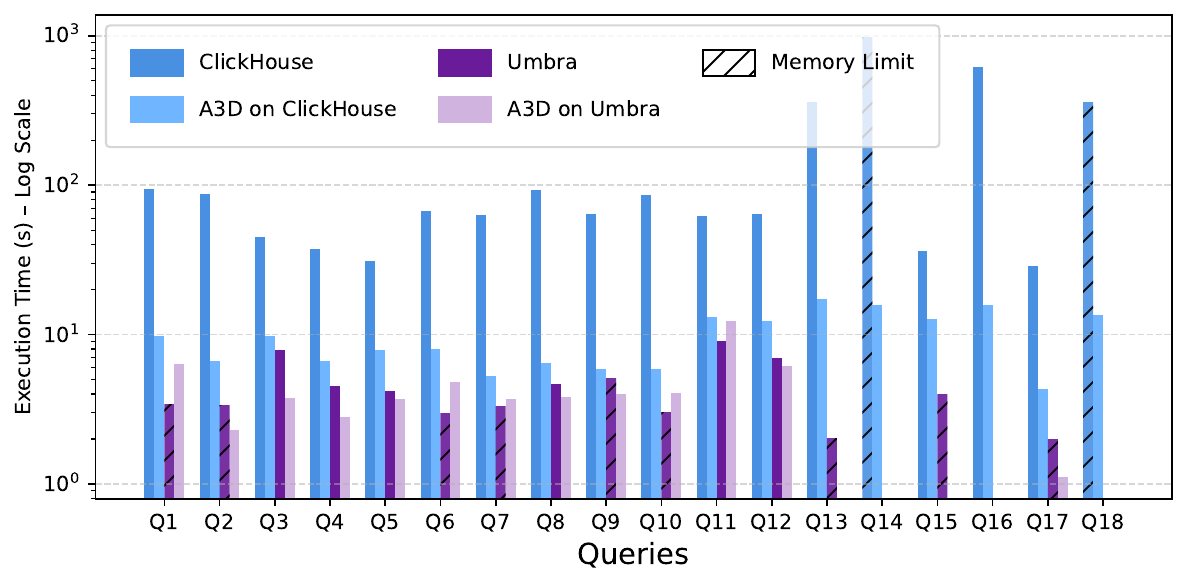}\vsqueezeabovecaption\caption{ClickHouse and Umbra runtime: native vs. A3D-RA.}
    \label{fig:OpenseeQueries}
\end{figure}

\begin{figure}[t]
    \centering
        \includegraphics[width=0.45\textwidth]{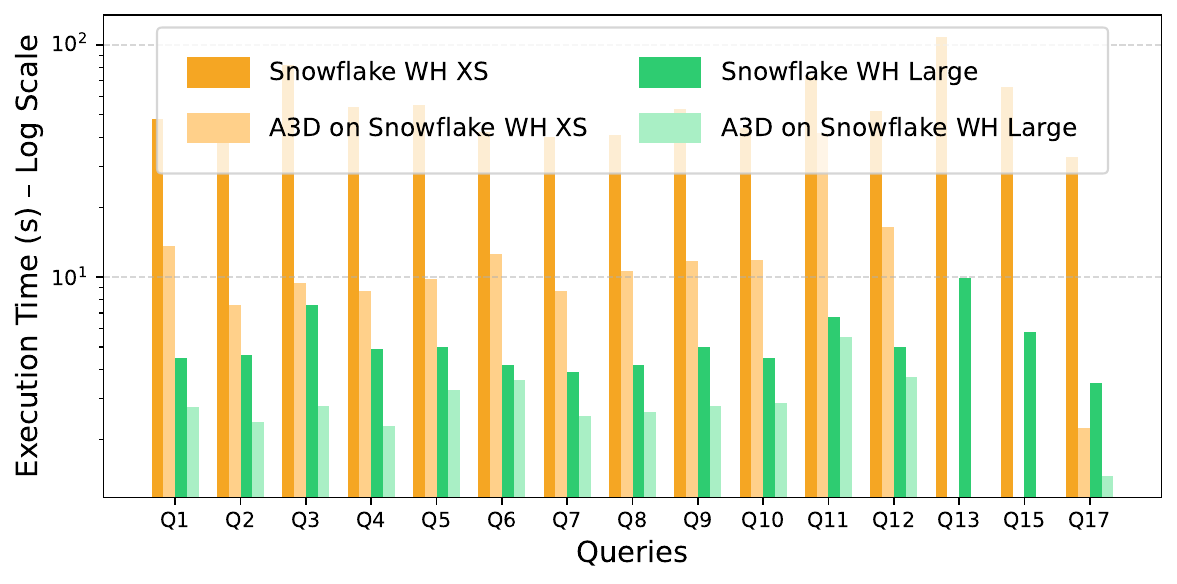}\vsqueezeabovecaption\caption{Snowflake runtime: native vs. A3D-RA.}
    \label{fig:sw}
    \vsqueezeaftercaption{}
\end{figure}

Figure~\ref{fig:OpenseeQueries} presents the execution time comparison with ClickHouse and Umbra, and Figure~\ref{fig:sw} shows the results for Snowflake on two warehouse configurations (X-Small and Large).
In both figures, hatched bars indicate queries that encounter memory limit errors, and missing bars represent queries that could not be translated due to unsupported language features.

\paragraph{ClickHouse.}
The A3D-Optimizer delivers significant performance improvements across all successfully translated queries.
For 16 queries that execute successfully in native ClickHouse, A3D achieves speedups ranging from $2.84\times$ to $38.75\times$, with a mean of $11.02\times$.
The most remarkable improvements are observed for Q16 ($38.75\times$ faster, from 614.6s to 15.9s), Q13 ($20.85\times$ faster), and Q8 ($14.30\times$ faster).
Notably, queries Q14 and Q18 encounter memory limit errors in native ClickHouse but execute successfully with A3D optimization (completing in 15.8s and 13.5s respectively), demonstrating that our transformations not only improve performance but also enable the execution of otherwise infeasible queries by reducing intermediate result sizes.

\paragraph{Umbra.}
For Umbra, A3D optimization demonstrates particularly strong results in resolving memory limitations.
Seven queries (Q1, Q2, Q6, Q7, Q9, Q10, Q17) that fail with memory limit errors in native Umbra execute successfully with A3D optimization, completing in 1.1–6.4s.
This highlights the effectiveness of our transformations in reducing memory consumption through early filtering and pre-aggregation.
For six queries that execute successfully in native Umbra (Q3, Q4, Q5, Q8, Q11, Q12), A3D achieves moderate speedups with a mean of $1.32\times$, demonstrating consistent but modest performance gains.
Five queries (Q13, Q14, Q15, Q16, Q18) could not be translated by A3D due to unsupported language features.

\paragraph{Snowflake.}
On the X-Small (XS) warehouse, A3D achieves consistent speedups for 13 evaluated queries (Q1–Q12, Q17), ranging from $1.76\times$ to $14.68\times$ with a mean of $5.35\times$.
The most significant gains are observed for Q17 ($14.68\times$), Q3 ($8.70\times$), and Q4 ($6.19\times$).

On the Large warehouse, baseline Snowflake performance improves substantially due to increased computational resources.
A3D optimization continues to provide speedups for all 13 evaluated queries, ranging from $1.17\times$ to $2.73\times$ with a mean of $1.75\times$.
The best improvements are for Q3 ($2.73\times$), Q17 ($2.50\times$), and Q4 ($2.15\times$).

Results show that A3D optimization delivers significant performance gains across different system architectures, with particularly strong results on ClickHouse (mean $11.02\times$) and Snowflake XS (mean $5.35\times$). 
In addition to improving raw performance, A3D can also enable the execution of queries that would otherwise exceed memory limits. For example, queries 14 and 18 are not feasible on ClickHouse without A3D optimizations.

\subsubsection{Rule Impact Analysis}
\label{sec:rule-impact-analysis}

To isolate the source of the observed gains, we evaluate whether they stem from a single dominant rule or the combined effect of the entire set using synthetic workloads on ClickHouse, which supports all queries.

\begin{figure}
    \centering
    \includegraphics[width=0.4\textwidth]{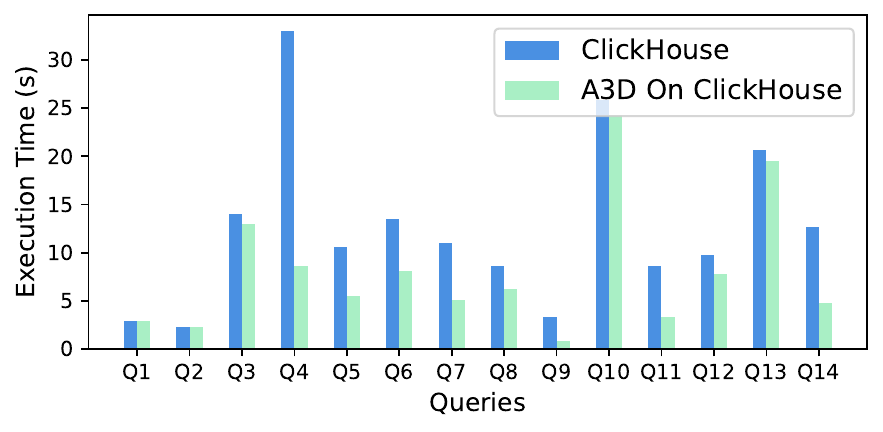}\vsqueezeabovecaption\caption{Impact of pushing filters under \texttt{ArrayJoin}.} 
    \label{fig:prefilters}
     \vsqueezeaftercaption{}
\end{figure}

\paragraph{Impact of pushing down \texttt{filters} under \texttt{ArrayJoin}.}
Fig.~\ref{fig:prefilters} illustrates rules (\ref{R2.1}–\ref{R2.3}), which push filters below \texttt{ArrayJoin}. A3D-Optimizer achieves speedups ranging from 1.7× to 4× by jointly reducing cardinalities vertically (through $\sigma$ filters) and horizontally (through \texttt{arrayFilter}). 
For Q4, horizontal filtering yields a 3.8× speedup: even with moderately selective filters ($\approx$50\%), horizontal reduction drastically limits flattening costs on large arrays.
For other queries (Q5–Q7, Q11, Q14), gains range from 1.7× to 2.6× due to combined vertical and horizontal filtering.
Queries with minimal improvement (Q1–Q3, Q10, Q13) have low filter selectivity, limiting optimization opportunities.

\begin{figure}
    \centering
    \includegraphics[width=0.4\textwidth]{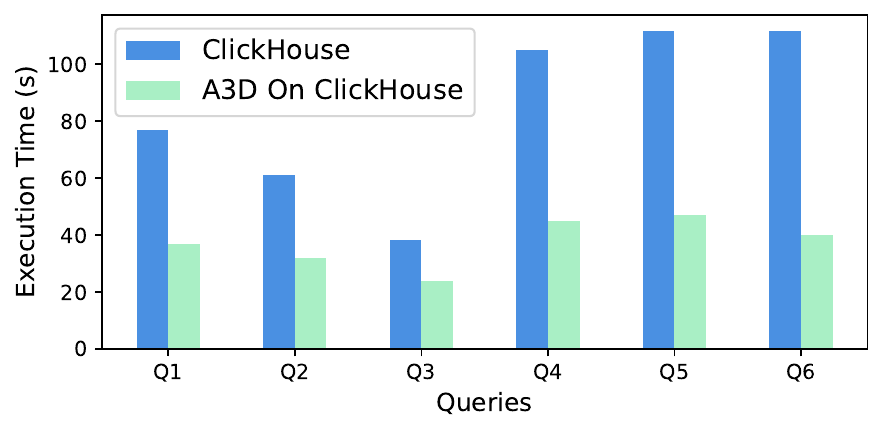}
   \vsqueezeabovecaption\caption{Impact of pushing \texttt{derive} under \texttt{ArrayJoin}.}
    \label{fig:derive vs arrayJoin}
\end{figure}

\paragraph{Impact of pushing down \texttt{derive} under \texttt{ArrayJoin}.}
Fig.~\ref{fig:derive vs arrayJoin} shows rules (\ref{R5.1}, \ref{R5.2}), which push \texttt{derive} beneath \texttt{ArrayJoin}. A3D-Optimizer achieves speedups ranging from 1.61× to 2.79× and consistently outperforms ClickHouse's native optimizer, where \texttt{derive} operations remain above array flattening.
For Q1–Q2, precomputing independent derivations avoids redundant evaluations on duplicated rows.
For Q3–Q6, applying transformations at the array level before flattening avoids per-row function calls. 

\begin{figure}
    \centering
    \includegraphics[width=0.4\textwidth]{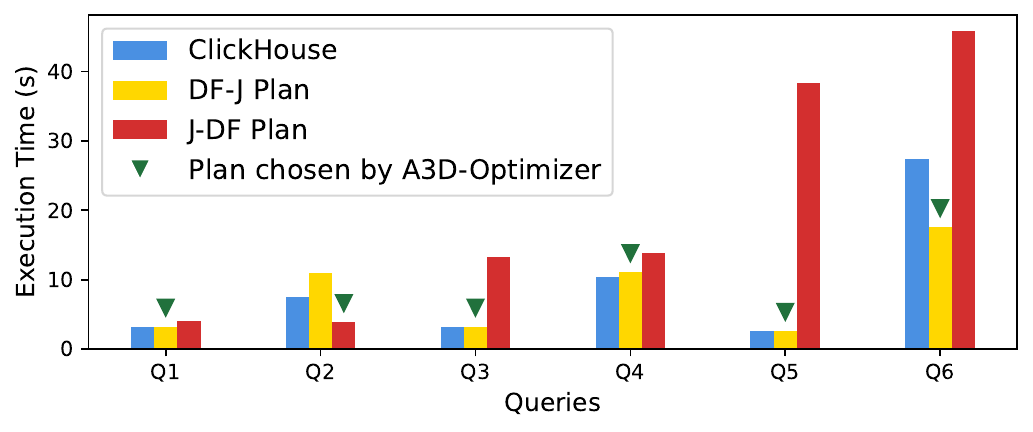}
  \vsqueezeabovecaption
  \caption{Non-invertible filters: DF-J and J-DF plans.} 
    \label{fig:DF vs join}
 \vsqueezeaftercaption{}
\end{figure}

\begin{figure}
    \centering
    \includegraphics[width=0.4\textwidth]{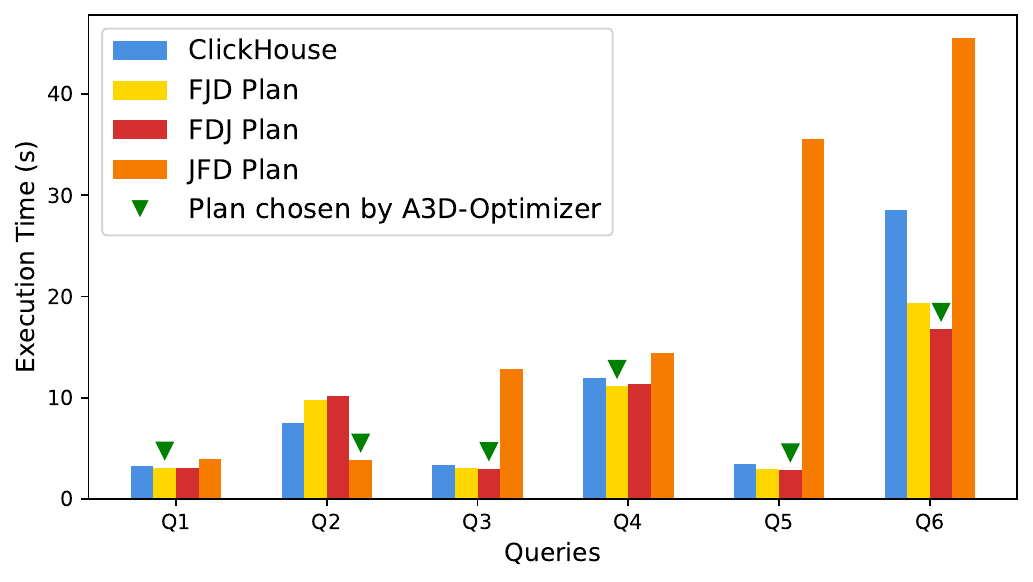}
    \vsqueezeabovecaption\caption{Invertible filters: alternative plans.
} 
%
    \label{fig:invertible filters}
    \vsqueezeaftercaption{}
\end{figure}

\paragraph{Invertibility of Filters.}  
Fig.~\ref{fig:DF vs join} and~\ref{fig:invertible filters} analyze the impact of filter invertibility on query performance.  
In the first case (Fig.~\ref{fig:DF vs join}), where the filter is non-invertible, two alternative plans are possible: applying the \texttt{derive} before the filter then join (DF-J), or performing the \texttt{join} first (J-DF).     
In contrast, Fig.~\ref{fig:invertible filters} illustrates the case of invertible filters, where three alternative plans can be considered: FJD, FDJ, and JFD.  %
%
The A3D-Optimizer consistently selects the most efficient plan.  
A notable performance gap is observed among the different strategies, particularly for Q5 and Q6, where the join operation significantly increases the cardinality of intermediate results.  
%

\begin{figure}[h]
    \centering
    \includegraphics[width=0.4\textwidth]
        {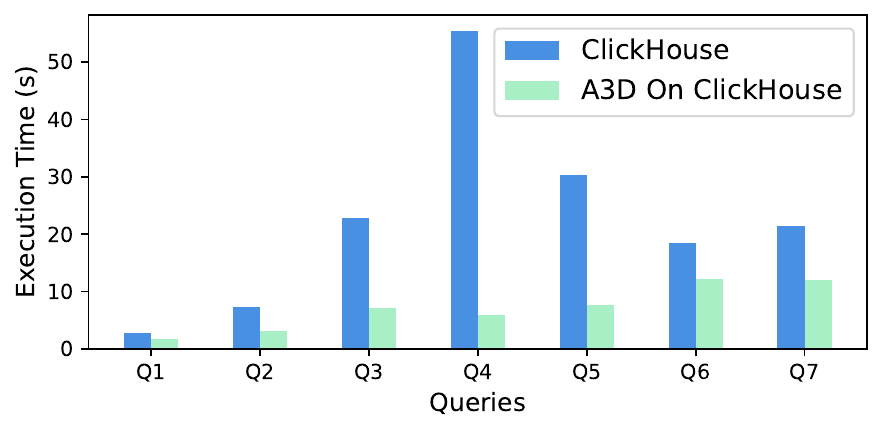}
  \vsqueezeabovecaption
  \caption{Impact of Introducing Pre-aggregations.} 
    \label{fig:preagg}
\vsqueezeaftercaption{}
\end{figure}
\paragraph{Introducing Pre-aggregations.}  
Fig.~\ref{fig:preagg} illustrates the impact of distributing aggregations through the introduction of pre-aggregations.
Performance improvements range from 1.52× to 9.22×.
The most significant gain is observed for Q4 (9.22× speedup), where pushing an aggregation below an \texttt{arrayJoin} on a large table substantially reduces intermediate data size.

Overall, the performance gains arise from a cumulative effect of all transformation rules, with a particularly strong impact from those pushing operators below \texttt{arrayJoin} to enable array-level computation, and from pre-aggregation rules—especially under \texttt{arrayJoin}—that reduce intermediate results and execution costs.

\subsubsection{Optimization Time}
\label{sec:optimization-time}

\begin{figure}[t]
    \centering
    \begin{tikzpicture}[
    node distance=0.3cm and 3cm,
    every node/.style={font=\footnotesize}
]

\node at (-2,0) (title_a) {(a) Pattern A};
\node[below=0.1cm of title_a] (sel1) {$\sigma_{\theta_y}$};
\node[below=of sel1] (der1) {$\delta_{y=f(a)}$};
\node[below=of der1] (aj1) {$\mu_{a}$};
\node[below=of aj1] (rel1) {$R$};

\draw[-latex] (rel1) -- (aj1);
\draw[-latex] (aj1) -- (der1);
\draw[-latex] (der1) -- (sel1);

\node[draw, dashed, rounded corners, fit=(sel1)(der1)(aj1), inner sep=2pt] {};

\node at (2,-0.4) (title_b) {(b) Pattern B};
\node[below=0.1cm of title_b] (sel2) {$\sigma_{\theta_1 \land ... \land \theta_n}$};
\node[below=of sel2] (aj2) {$\mu_{arr_1,\ldots,arr_n}$};
\node[below=of aj2] (rel2) {$R$};

\draw[-latex] (rel2) -- (aj2);
\draw[-latex] (aj2) -- (sel2);

\node[draw, dashed, rounded corners, fit=(sel2)(aj2), inner sep=2pt] {};

\end{tikzpicture}
    \vsqueezeabovecaption
    \caption{Two example patterns used in scalability analysis.}
    \label{fig:algebraic_plan_example}
   \vsqueezeaftercaption{}
\end{figure}
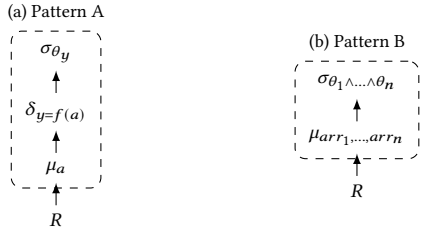

\begin{figure}[t]
    \centering
    \includegraphics[width=0.45\textwidth]{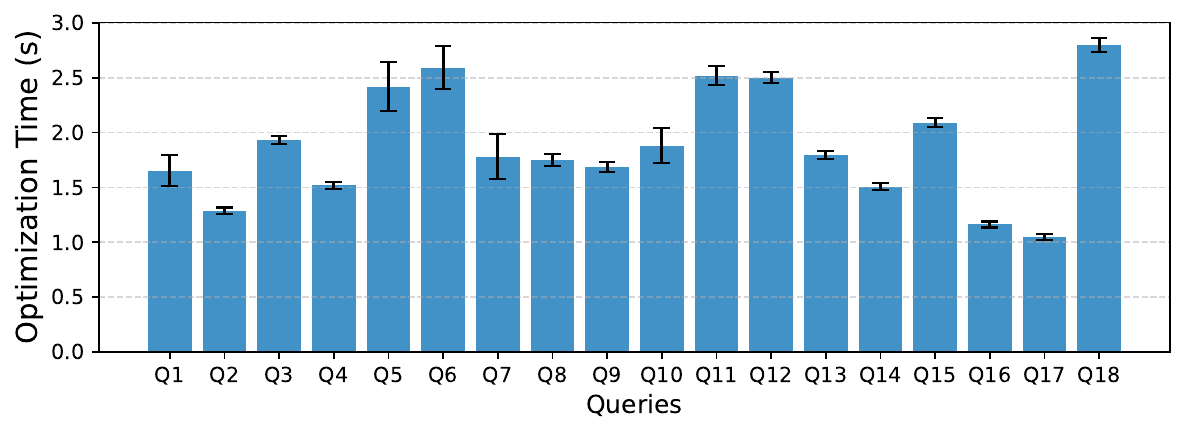}
  \vsqueezeabovecaption
  \caption{Optimization time for 18 real-world queries.}
    \label{fig:optimization_time}
\vsqueezeaftercaption{}
\end{figure}

\begin{figure*}[t]
    \centering
    \includegraphics[trim=0 0.3cm 0 0, clip,width=\textwidth, height=0.16\textheight]{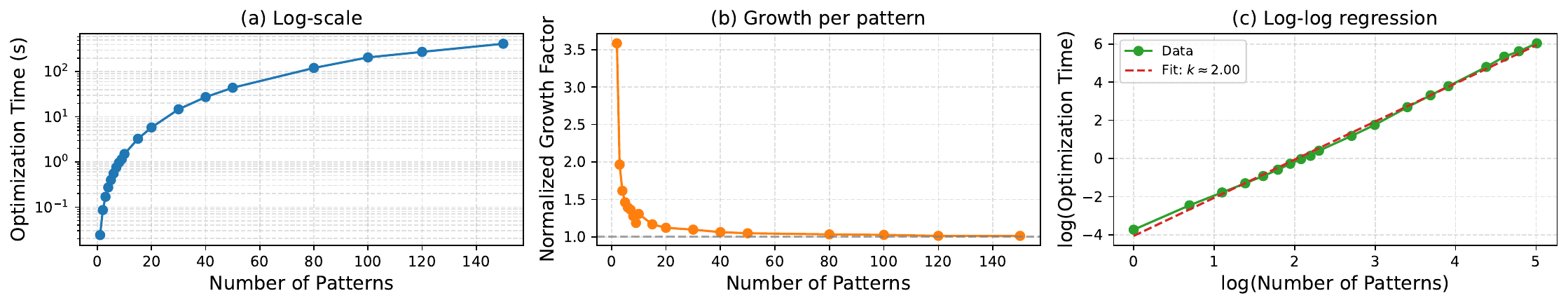}
  \vsqueezeabovecaption
   \vsqueezeabovecaption
  \caption{Scaling optimization behavior with number of patterns: (a) log-scale, (b) growth factor, (c) log-log regression.}
    \label{fig:scaling_patterns}
    \vsqueezeaftercaption{}
\end{figure*}

\begin{figure*}[t]
    \centering
    \includegraphics[trim=0 0.3cm 0 0, clip,width=\textwidth]{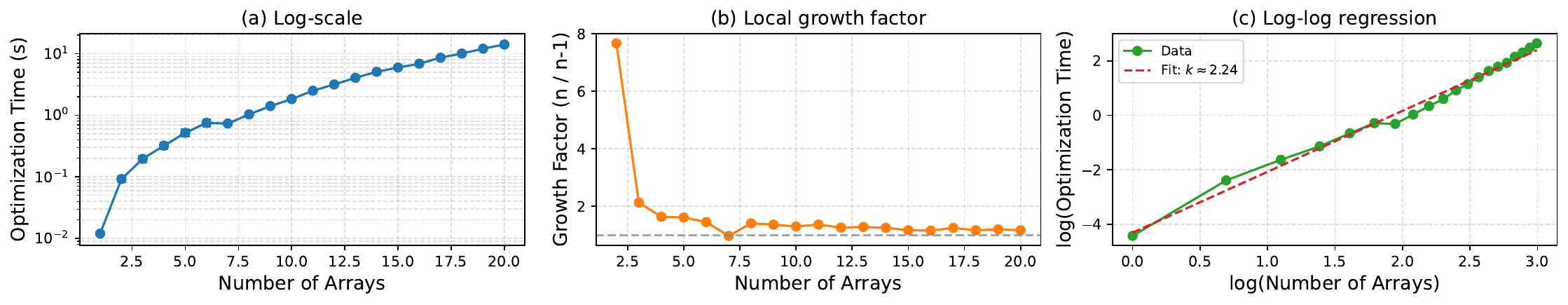}
  \vsqueezeabovecaption
  \vsqueezeabovecaption
  \caption{Scaling optimization behavior with number of arrays per pattern: (a) log-scale, (b) growth factor, (c) log-log regression.}
    \label{fig:scaling_arrays}
    \vsqueezeaftercaption{}
\end{figure*}

We now evaluate the optimization overhead introduced by the A3D-Optimizer. 
Optimization time is measured from the input query to the optimized algebraic plan—before translation to the target DBMS (Figure~\ref{fig:archi}). This measurement is thus independent of the underlying database system.

\paragraph{Optimization Time on Real-World Queries.}
Figure~\ref{fig:optimization_time} shows the optimization time for the 18 real-world queries.
Optimization times range from 1.05s to 2.80s, with a mean of 1.89s.
To assess the cost-benefit trade-off, we compute the optimization payoff (OP) as the ratio of execution time gain to optimization cost for each query across all evaluated systems.
For ClickHouse, the OP ranges from 9.6× to 515.7× with a mean of 70.4×, demonstrating that each second of optimization saves 70.4 seconds of execution time on average.
For instance, query Q16 incurs 1.16s of optimization cost while achieving 598.7s of execution time savings, resulting in an OP of 515.7×.
For Snowflake XS, the OP ranges from 11.3× to 37.6× with a mean of 21.5×.
For Umbra and Snowflake Large, the OP is more modest (0.8× and 1.1× on average respectively), as most Umbra queries encounter memory limits in the baseline and Snowflake Large already achieves very high performance, limiting the absolute time savings.
Overall, the optimization overhead is negligible compared to the performance benefits obtained, especially on ClickHouse and Snowflake XS where OP exceeds 20× on average.

\paragraph{Scalability Analysis.}
We assess how optimization time scales with query complexity, which we define as the number of \emph{patterns} present in a query.
A \emph{pattern} is a sequence of algebraic operators applied to array-typed columns (see Figure~\ref{fig:algebraic_plan_example}).
As shown in Section~\ref{sec:complexity}, A3DRA optimization admits a theoretical polynomial upper bound for non-join array operations, and adding patterns directly increases this structural complexity. %
To study this, we vary two parameters: the number of patterns and the number of arrays per pattern. We consider two patterns shown in Figure~\ref{fig:algebraic_plan_example}.
Pattern A represents a typical transformation pipeline with ArrayJoin, Derive, and Filter operators, while Pattern B shows multiple arrays being processed with combined filters.
We use Pattern A to evaluate scalability with respect to the number of patterns (Figure~\ref{fig:scaling_patterns}) and Pattern B to evaluate scalability with respect to the number of arrays per pattern (Figure~\ref{fig:scaling_arrays}).

Figure~\ref{fig:scaling_patterns} presents the scalability analysis with respect to the number of patterns (pattern A).
As the number of patterns increases from 1 to 150, optimization time grows from 0.024s to 416s.
The log-log regression analysis (subplot c) reveals a polynomial complexity with an exponent $k \approx 2.0$, indicating quadratic growth ($R^2>99\%$).
The normalized growth factor (subplot b) stabilizes around 1.1–1.2 per additional pattern, suggesting predictable scaling behavior.

Figure~\ref{fig:scaling_arrays} examines scalability with respect to the number of arrays per pattern (pattern B).
When varying the number of arrays from 1 to 20, optimization time increases from 0.012s to 14s.
The log-log regression yields $k \approx 2.2$, indicating super-quadratic but still polynomial growth ($R^2 = 98.84\%$).
The local growth factor (subplot b) shows an initial spike but stabilizes around 1.15 for higher array counts.
This suggests that while multiple arrays increase optimization complexity, the cost remains manageable for typical analytical queries that rarely exceed 10–15 array columns per pattern.

\section{Related Work}

\paragraph{Arrays in Databases}
The idea of supporting arrays in database systems has a long history. Early work on the nested relational model and the Non First Normal Form (NF$^2$) data model \cite{jaeschke-pods82,ozsoyouglu-tods87} extended the relational paradigm to allow attributes that are themselves collections. %
At the query language level, several algebras and formalisms for arrays were proposed, 
such as AQL~\cite{libkin-sigmod96}, RasQL~\cite{baumann-sigmod98}, AML~\cite{marathe-vldbj2002}, RAM~\cite{ballegooij-edbt04}, and ArrayQL \cite{ArrayQL12,neumann-edbt22}. %
Building on these foundations, several dedicated array database systems were developed. RasDaMan~\cite{baumann-sigmod98} pioneered array-oriented storage and query processing, emphasizing raster data. SciDB~\cite{stonebraker-pvldb09} introduced an array-native data model with specialized operators for scientific and analytical workloads. TileDB~\cite{papadopoulos-pvldb16} proposed a universal storage engine for dense and sparse arrays. MonetDB~\cite{monetdb2005,monetdb2012} sought to integrate array processing into a columnar relational system, proposing an SQL dialect with array operators. %
At the same time, mainstream relational systems have gradually incorporated arrays, though typically in an ad-hoc fashion. PostgreSQL introduced native array types and functions, enabling applications to store and manipulate arrays inside relations. More recent systems such as Google BigQuery~\cite{dremel-bigquery-pvldb10}, SparkSQL~\cite{sparkSQL-sigmod15}, Snowflake~\cite{snowflake2016}, DuckDB (in-memory)~\cite{duckdb-sigmod2019}, Umbra~\cite{neumann-umbra2020}, and most recently ClickHouse~\cite{schulze2024clickhouse} extend SQL with array-valued functions and user-defined operators. The work on integrating ArrayQL in Umbra~\cite{neumann-ssdbm21,neumann-edbt22} provides a set of array operators expressible in relational algebra, its goal is to embed multidimensional array computations into the relational engine. In contrast, our work extends the relational algebra itself with array-valued attributes and compositional transformation rules, enabling systematic optimization of queries that freely interleave relational and array operators.

A recent survey \cite{rusu-ftdb2023} provides an in-depth comparison of existing array data management techniques. It observes that no array algebra and query language have gained general acceptance so far. A key reason is that most existing approaches rely heavily on user-defined functions or system-specific extensions, without providing an explicit algebraic foundation that integrates with relational optimization. Our work takes a different approach: instead of building a specialized array DBMS or exposing arrays only via ad hoc extensions, we extend relational algebra itself to treat arrays as first-class citizens, in the spirit of the seminal NF$^2$ work~\cite{jaeschke-pods82}. 
We instantiate the framework on top of ClickHouse~\cite{schulze2024clickhouse}, Umbra~\cite{neumann-umbra2020} and Snowflake~\cite{snowflake2016}, demonstrating its practical benefits.  

\paragraph{Denormalized Data and Column Stores}
Beyond arrays, the database community has long studied richer data models that extend relations with nested or semi-structured data. The nested relational algebra \cite{jaeschke-pods82} formalized operators over nested collections, and subsequent work on unnesting and query flattening \cite{buneman-tcs95, fegaras-sigmod98} addressed optimization challenges. Semi-structured data in JSON format has driven widespread extensions to SQL. 
These extensions illustrate a general trend: relational systems increasingly embrace denormalized data, moving beyond traditional first normal form. %
This trend is particularly pronounced in analytical workloads, where denormalized schemas with wide tables and nested attributes are common. Column-oriented database systems, beginning with MonetDB \cite{monetdb2005} and C-Store \cite{stonebraker-vldb05} and followed by many successors, have proven to be a natural fit for such workloads. They exploit contiguity, compression, and vectorized execution. ClickHouse \cite{schulze2024clickhouse} is a recent column store, with native support for array-valued attributes, which was shown to outperform several state-of-the-art systems \cite{schulze2024clickhouse}, including PostgreSQL, Redshift~\cite{redshift-sigmod2015}, Pinot~\cite{pinot-sigmod18}, Umbra \cite{neumann-umbra2020} and Snowflake~\cite{snowflake2016}. Our prototype implementation,  instanciated on top of ClickHouse~\cite{schulze2024clickhouse}, Umbra~\cite{neumann-umbra2020}, and Snowflake~\cite{snowflake2016}, demonstrates that algebraic extensions can still unlock further optimization opportunities, with each of these systems.

\section{Conclusion}
This paper introduced an extended relational algebra supporting array-valued attributes, together with a framework for algebraic reasoning and optimization. We defined its formal foundations, a complete set of equivalence-preserving transformation rules, and a plan enumeration strategy with an optimality guarantee and a polynomial complexity in all non-join operators. 
We designed A3D-RA as a modular, backend-independent optimization layer that can be instantiated over existing analytical database systems. Experimental results across three analytical engines on a real-world workload demonstrate that the framework can improve query execution performance without requiring modifications to the underlying execution engines. These results show the benefits of treating array operations as first-class algebraic constructs, allowing the optimizer to perform global, systematic rewrites that jointly consider relational and array operators. 

\clearpage

\balance

\bibliographystyle{ACM-Reference-Format}
\bibliography{refs}


\end{document}